\documentclass[11pt]{article}
\usepackage{booktabs}
\usepackage{mathrsfs}
\usepackage{amsmath}
\usepackage{mathrsfs}
\usepackage{amssymb}
\usepackage{color}
\usepackage{psfrag}
\usepackage{multirow}
\usepackage{graphicx}
\usepackage{subfigure}
\usepackage{rotating}
\usepackage{cite}
\usepackage[colorlinks=true,linktocpage=true,linkcolor=blue,citecolor=blue]{hyperref}
\usepackage[section]{placeins}
\usepackage{amsfonts}
\usepackage{ifpdf}
\ifpdf
\else

\renewcommand{\bf}{\textbf}

\renewcommand{\it}{\textit}
\newcommand{\mc}{\mathcal}

\newcommand{\mb}{\mathbf}

\newcommand{\be}{\begin{equation}}
\newcommand{\ee}{\end{equation}}
\newcommand{\bes}{\begin{subequations}}
\newcommand{\ees}{\end{subequations}}
\newcommand{\ben}{\begin{equation*}}
\newcommand{\een}{\end{equation*}}
\newcommand{\bea}{\begin{eqnarray}}
\newcommand{\eea}{\end{eqnarray}}
\newcommand{\h}{\frac{1}{2}}
\newcommand{\p}{\partial}
\newcommand{\tin}{\tan^{-1}}

\renewcommand{\d}{\delta}
\newcommand{\D}{\Delta}

\renewcommand{\L}{\Lambda}

\renewcommand{\t}{\theta}

\newcommand{\vsp}[1]{\vspace{#1 mm}}
\begin{document}
\topmargin -24pt
\oddsidemargin 0mm

\vspace{2mm}

\begin{center}
{\Large \bf {From Classical Periodic Orbits in Integrable $\pi$-Rational Billiards to Quantum Energy Spectrum}}

\vsp{10}
Subhasis Panda$^{a}$ \footnote{E-mail: subhasis@phy.nits.ac.in},
Sabyasachi Maulik$^{b,c}$ \footnote{E-mail: sabyasachi.maulik@saha.ac.in},
Somdeb Chakraborty{$^{d}$ \footnote{E-mail: somdeb.ch.12@gmail.com}} and
S. Pratik Khastgir$^{b,e}$ \footnote{E-mail: pratik@phy.iitkgp.ac.in} 

\vspace{4mm}

{\em
$^{a}$ Department of Physics, N.I.T. Silchar, 788010, Cachar, India \\
$^{b}$ Department of Physics, IIT Kharagpur, Kharagpur 721302, India \\ 
$^{c}$ Saha Institute of Nuclear Physics, 1/AF Bidhannagar, Kolkata 700064, India  \\
$^{d}$ Department of Physics, City College, 102/1, Raja Rammohan Sarani, Kolkata 70009, India\\
$^{e}$ Centre for Theoretical Studies, IIT Kharagpur, Kharagpur 721302, India 
}
\end{center}
\vsp{10}
\begin{abstract}
In the present note, we uncover a remarkable connection between the length of periodic orbit of a classical particle enclosed in a class of 2-dimensional planar billiards and the energy of a quantum particle confined to move in an identical region with infinitely high potential wall on the boundary. We observe that the quantum energy spectrum of the particle is in exact one-to-one correspondence with the spectrum of the amplitude squares of the periodic orbits of a classical particle for the class of integrable billiards considered. We have established the results by geometric constructions and exploiting the method of reflective tiling and folding of classical trajectories. 
We have further extended the method to $3$-dimensional billiards for which exact analytical results are scarcely available - exploiting the geometric construction, we determine the exact energy spectra of two new tetrahedral domains which we believe are integrable. We test the veracity of our results by comparing them with numerical results.
\end{abstract}

\section{Introduction}

The billiard problem, which draws inspiration from the game of billiards, has fascinated physicists for a long time - both in its classical and quantum versions. The dynamics of a billiard is controlled by the shape of its boundary and encapsulates all the features of a Hamiltonian system - ranging from completely regular to fully chaotic. Classical billiards are used to model a vast array of phenomena in realms of optics, acoustics and classical mechanics. Interestingly, the first billiard model can be traced back to the hard ball model of gas credited to Boltzmann and Lorentz \cite{Lorentz}. The quantum billiard, on the other hand, invites us to explore the exciting and rich field of quantum chaos. Besides being theoretically appealing, quantum billiards also find application in frontier areas like nanotechnology and mesoscale physics. For instance, electrons in nanodevices, like in quantum dots, are often restricted to two spatial dimensions, and the quantum billiard can efficiently capture the salient features of the dynamics. Lately, quantum billiards have also been recreated in the laboratory by means of different techniques including electrical resonance circuits \cite{Bengtsson} and microwave cavities \cite{Kudroli}. \\
In its classical variant, we define a billiard table as a Riemannian manifold, $\mc{M}$, with a piecewise smooth, rigid boundary, $\p\mc{M}$, and study the trajectory of the billiard ball - a point particle of mass $m$, in $\mc{M}$. The particle's dynamics follows from the Hamiltonian
\be 
H(\mb{x}, \mb{p}) = \frac{1}{2 m} g_{ij} p^i p^j + V(\mb{x}),
\ee
where $p^i$ are the momenta, $g_{ij}(\mb{x})$ is the metric tensor at $\mb{x} \in \mc{M}$ and the potential energy, $V(\mb{x})$, is defined by
\be 
V(\mb{x}) = \begin{cases}
 0 & \mb{x} \in \mc{M} \\
 \infty & \mb{x} \notin \mc{M}.
 \end{cases}
\ee
The specific choice of $V(\mb{x})$ guarantees that the particle moves along a geodesic in $\mc{M}$ interspersed by collisions on $\p\mc{M}$. The collisions are assumed to be energy-conserving and specular. \\
In the present note, we shall take $\mc{M}$ to be a patch of $2$-dimensional Euclidean space, $\mathbb{E}^2$, and accordingly,
identify $g_{ij}$ with $\d_{ij}$. Planar billiards have long acted as a testing ground for the quantum-classical correspondence since 
they can easily be adapted to explore a rich variety of dynamical features just by tuning the geometry \cite{Brack, Chen, Chen1, 
Doncheski, Lin, Liu, Chen2, Wright, Altmann}. Most of the activities have focused on non-integrable and irregular billiards and 
investigating the connection between classical periodic orbits and quantum eigenstates and energy spectrum \cite{Altmann, Heller, 
McDonald, Tomosovic, Zaslavsky, Gutzwiller, Sridhar, Tomosovic1, Tomosovic2, Berry}. On the other hand, periodic orbits are profusely found in integrable billiards with symmetric shapes \cite{Brack, Gutzwiller, Robinett, Robinett1, Robinett2, Doncheski1, Styer, Robinett3, Robinett4, Wright1, Robinett5, Lu, Matzkin, Fonte, Robinett6, Macek, Lee, Robinett7}. Planar billiards can roughly be classified into three types depending upon the choice of $\p \mc{M}$: elliptic (convex billiards), hyperbolic (concave billiards), or parabolic (polygonal billiards). Here our interest lies only in the last class. A variety of dynamical systems can be mapped to polygonal billiards - for example, two elastic, massive particles confined to move on a $1$-dimensional segment is isomorphic to a right-triangle billiard whose dimensions depend on the ratio of the masses \cite{ MasurTab}. Similarly, a system of three hard rods, sliding on a frictionless ring and making elastic collisions, is equivalent to an acute-triangle billiard \cite{Glashow}. In addition, if all angles of a polygonal billiard table are rational multiples of $\pi$, the billiard is called a rational billiard. In all the examples we discuss, our billiard will be $\pi$-rational. There are reasons why rational billiards are picked out for special treatment. First of all, because it is easy! Without the rationality assumption, there are very few tools available to tackle the problem. On the contrary, once the rationality assumption is enforced, a lot of things can be proven. Secondly, the assumption is not too restrictive in the sense that even with this constraint in place, it is possible to explore a rich and diverse set of phenomena in polygonal billiards. Thirdly, the rationality assumption enables us to make surprising and beautiful connections to algebraic geometry, Teichm\"{u}ller theory, ergodic theory on homogeneous spaces, and other areas of mathematics. Here we are interested in rational billiards for a different reason - in $\pi$-rational billiards, it is possible to find orbits that are periodic. 

The next section describes rational billiards in detail. Various subsections are individually dedicated to treating different types of integrable $\pi$-rational billiards in two dimensions. For each of the integrable domains, the correspondence between the quantum eigenspectrum and classical periodic trajectory amplitudes is established. Section 3 deals with billiards in three dimensions. We find out the exact eigenspectra for two new tetrahedral domains. We also show the geometric constructions for the corresponding domains. In section 4, we summarize our findings and conclude.

\section{Rational billiard}

We consider a rational polygon with $s$ sides for which the vertex angles are given by $\frac{m_i}{n_i}\pi,~i=1,2,\cdots,s$ and $m_i$
and $n_i$ are integers. Trajectories in this billiard can take at most $2N$ different directions after all successive collisions, where $N$ is the least common multiple of the $n_i$'s. The invariant manifold, ${\cal I}$, of this billiard is made with $2N$ copies of the original polygon by successive reflections. Identifying the opposite sides of ${\cal I}$, one constructs a genus $g$ surface \cite{Richensberry,Gutkin,Zyczko}, where, \be
g=1+\frac{N}{2}\sum_{i=1}^{s}\frac{m_i-1}{n_i}. \label{genus}
\ee
For integrable billiards, genus of the invariant manifold is unity. A simple inspection of (\ref{genus}) shows that a rectangle is integrable.
The only other integrable billiards are the three triangles with vertex angles $(\pi/3, \pi/3, \pi/3)$, $(\pi/2,\pi/4,\pi/4)$ and $(\pi/2,\pi/3,\pi/6)$. Billiards corresponding to genus 2 invariant manifolds are termed pseudo-integrable - here one finds the partial spectrum of the problem to be exact. For instance, a rhombus with angles $(\pi/3, 2\pi/3,\pi/3,2\pi/3)$ is pseudo-integrable \cite{Jain,Shudo}.
To put it simply, a shape which can fill $\mathbb{E}^2$ by reflective replication without leaving holes, would be integrable \cite{Richensberry}. However, translational tiling does not guarantee integrability - that is why, a rhombus is not integrable. So for reflective tiling with a polygon, necessarily an even number of self-similar tiles should meet at each vertex. Hence, the vertex angles
of integrable polygons are restricted to $\pi/2,\pi/3,\pi/4$ and $\pi/6$ only. Further, as mentioned earlier, we shall limit ourselves to only those $\pi$-rational billiards which are integrable ($g=1$). 

We wish to explore the connection between the amplitude of a classical periodic orbit in an integrable $\pi$-rational billiard and the spectrum of the Helmholtz operator defined on a similar domain. Various studies have been undertaken in this direction in the past, notably using the periodic orbit theory and the semi-classical behavior of wave packets \cite{Brack, Balian, Berry2}. Here we wish to address the same problem using a method of geometric construction. To set the stage for discussion, let us first classify the various types of orbits. A terminal orbit is one which hits a vertex on $\p \mc{M}$ and gets annihilated. A $p$-periodic orbit is one which retraces its path after suffering $p$ collisions on $\p \mc{M}$. Finally, an infinite orbit is one that is neither terminal nor periodic. In this note, we shall only consider periodic orbits. We define the amplitude as the maximum distance of the particle from its initial shooting point along the trajectory. \\

\subsection{The square billiard}

\begin{figure} [ht] 
\begin{center} 
\includegraphics[scale=0.7, angle=0]{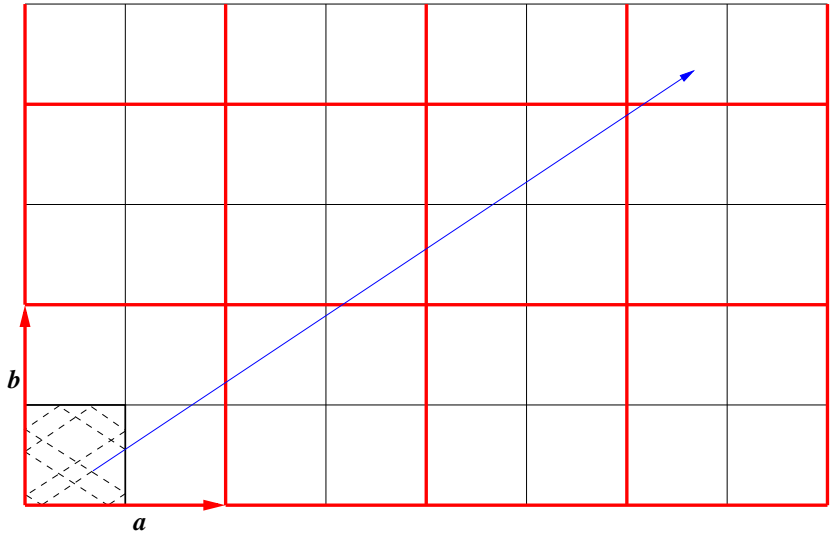} 
\caption{{\label{sqlatt}} Construction of a periodic orbit in a square billiard with $m=3$ and $n=2$.}
\end{center}
\end{figure}

\begin{figure} [ht] 
\begin{center} 
\includegraphics[scale=0.7, angle=0]{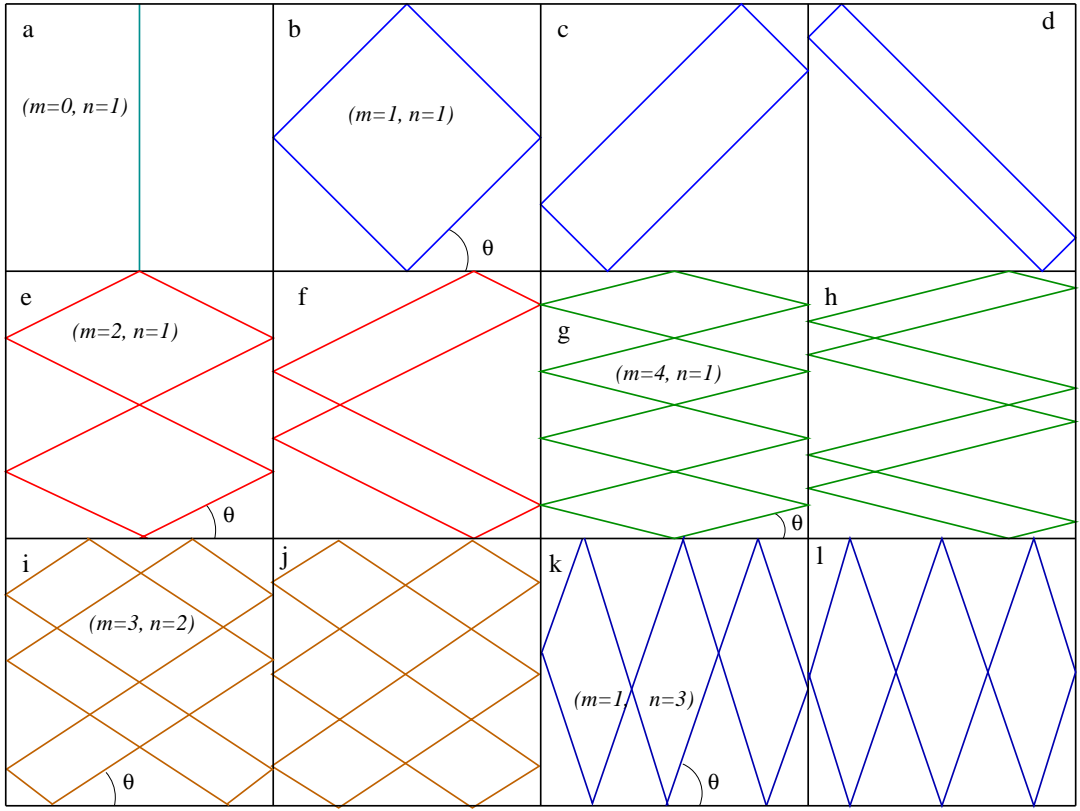} 
\caption{{\label{square}} Periodic trajectories for different shooting angles, $\t$, and different points of projection in a square billiard. The shooting angle is obtained from the relation $\tan \t = \frac{n}{m}$.}
\end{center}
\end{figure}

First of all, we consider a particle moving in a square billiard. The potential energy is of the form
\be \label{pot}
V(x,y) = \begin{cases}
 0 & 0\leq x, y\leq L\\
\infty & \text{otherwise},
 \end{cases}
\ee
where the left-bottom corner of the square is placed at the origin. Let the particle be projected from a point $(x_0, y_0)$ at an angle $\t$ with respect to the $x$-axis, i.e., $\t$ is measured counterclockwise from the $x$-axis. We shall throughout assume $0 \leq \t \leq \frac{\pi}{2}$. The case of negative $\t$ simply amounts to a mirror reflection about the $x$-axis. The particle suffers multiple collisions on the edges and finally comes back to $(x_0, y_0)$ again making an angle $\t$ to the $x$-axis. Such a motion is possible only when $\tan \t$ is of the form $\frac{n}{m}$, where $m$ and $n$ count the number of collisions made by the particle on the $y$ and the $x$-axes respectively, in the complete circuit. To find the amplitude of the particle, we shall make use of the \it{unfolding} trick. Instead of reflecting the path of the particle at a collision on the edge, we reflect the square along that particular edge and consider an extension of the original path of the particle in the reflected billiard. Subsequent bounces on the edges are treated in a similar manner. The unfolding trick can be employed subject to the fulfillment of two conditions. Firstly, by reflecting the billiard, it is possible to tile the whole of $\mathbb{R}^2$. Secondly, after the billiard has been reflected, the trajectory of the particle in the reflected domains should reduce to some well-known trajectory that is easier to handle. By reflection of the original domain, $\mc{M}$, we are thus led to a lattice where a unit cell is defined by $\mc{M}$. It is now clear that the trajectory of the particle in $\mc{M}$ will be $p$-periodic only if the original straight line passes through a point in the lattice whose coordinates are of the form $(x_0 + 2mL, y_0 + 2nL)$, with $m,n$ integers. Of course, when this straight line is folded back to $\mc{M}$, it reduces to the original trajectory which bounces off the edges at $2(m+n)$ points. Fig.\ref{sqlatt} shows one such trajectory, which starts off in the fundamental region at $(x_0=\frac{2}{3}L,y_0=\frac{1}{3}L)$ and ends at $(\frac{2}{3}L+6L, \frac{1}{3}L+4L)$. The folded trajectory in the fundamental region is shown with dashed lines. This trajectory is the same as one shown in Fig \ref{square}i. 
The case $m=n=0$ corresponds to a particle sitting idle at $(x_0, y_0)$. The case $m>0, n=0$ or, equivalently, $\tan \t = 0$ describes a particle traveling at $\t = 0$ and hitting an edge parallel to the $y$-axis. Likewise, the case $m=0, n>0$ or, $\tan \t \to \infty$ describes a particle moving at $\t = \frac{\pi}{2}$ and tracing a path parallel to the $y$-axis. Thus, we find that the case $m=0$, or, $n=0$ corresponds to the particle traveling tangentially to one of the axes and never colliding with the same and always colliding normally with other. These orbits do not generate closed loops and correspond to the Neumann spectrum of the problem. In the context of quantum mechanics, the wave function has to vanish on $\p \mc{M}$ which belongs to the Dirichlet spectrum. Henceforth, we shall restrict ourselves only to the case where $(m,n)$ are positive integers. It is worth noting at this stage that the time period or, the amplitude of the particle is insensitive to any change in the initial position of the particle $(x_0, y_0)$, as long as $\t$ is kept constant. Varying the initial point only yields a family of trajectories that are parallel to each other. Likewise, the transformation $\t \to \frac{\pi}{2} - \t$ interchanges $m$ and $n$, but leaves the amplitude unchanged. This is also evident from the fact that once a particle is shot at an angle $\t$, after the first reflection it bounces off at an angle $\frac{\pi}{2} - \t$ and for periodic orbits we could as well have started from this point. Let us now consider different pairs of $(m,n)$. Tuning these two parameters gives us a whole family of periodic orbits, each with different amplitude, $A(m,n)$ modulo the invariance under $(m,n) \to (n,m)$, as already discussed. This degeneracy in amplitude is, in fact, an expected one due to the $x \leftrightarrow y$ symmetry of the system. Apart from this, there are also accidental degeneracies where two different sets of ($m, n$) conspire to result in the same amplitude - for instance, $(m,n) = (1,7)$ and $(5,5)$ have the same amplitude. Another case of accidental degeneracy is furnished by the pairs, $(m,n) = (1,8)$ and $(4,7)$. We further define the quantity $\mc{A}(m,n) = A^2(m,n)$ and find out its value for some sample pairs of $(m,n)$. Let us start from the minimum amplitude. Clearly, this is realized for $(m,n) = (1,1)$ and accordingly $A(1,1)=\sqrt{2}L$ and $\mc{A}(1,1)=2L^2$. Such a situation arises when the projection angle is $\t = \frac{\pi}{4}$. The next member of the family is $A(1,2) = \sqrt{5}L$ and $\mc{A}(1,2)=5L^2$. We can proceed in a similar fashion and show $\mc{A}(2,2) = 8L^2, \mc{A}(1,3) = 10L^2, \mc{A}(2,3)=13L^2$ and so on.
In Fig.\ref{square}, a few representative trajectories are shown. Trajectories with the same values of $(m,n)$ but with different starting points are equivalent and shown in the same color. The first one corresponds to the case $m = 0, n = 1$ when the particle heads straight towards the opposite edge and, as discussed, is a part of the Neumann spectrum only. The next three cases are for $\t = \frac{\pi}{4}$ with $m = n = 1$ and correspond to different points of projection. However, in all the cases, $\mc{A}(1,1)$ remains the same. Details of the subsequent trajectories should be evident from Fig.\ref{square} itself. 

To set the stage for other geometries and for obtaining a generalized method of geometric construction of periodic orbits, we now adopt the following procedure. The original square lattice (shown in black in Fig.\ref{sqlatt}) is immersed in another square lattice with doubled lattice vectors, as shown in red in Fig.\ref{sqlatt}. Geometrically, the bigger lattice is created with two orthogonal basis vectors ${\bf {\it a}} = 2L{\bf i}$ and ${\bf{\it b}}=2L{\bf j}$, where ${\bf i}$ and ${\bf j}$ are unit vectors along the $x$ and the $y$-axes respectively. It is now clear that any vector, ${\bf V}$, of the form, $m{\bf {\it a}}+n{\bf{\it b}}$, can be folded as a periodic trajectory in the smaller immersed lattice, since this construction guarantees even number of horizontal and vertical crossings with the smaller lattice. For any arbitrary trajectory, the corresponding vector is ${\bf V}=m{\bf {\it a}}+n{\bf{\it b}}$ and
\be
A(m,n)=\frac{1}{2}|{\bf V}|
\ee
which leads to
\be
\mc{A}(m,n) =\frac{1}{4}{{\bf V}\cdot {\bf V}}= \left(m^2 + n^2 \right) L^2.
\ee

Let us now turn to the quantum side of the problem. We consider a particle of mass $\mu$ trapped in a $2$-dimensional square region of 
side $L$ by the potential defined in (\ref{pot}). The problem is solved by imposing proper Dirichlet condition, i.e., $\psi|_{\p \mc{M}}=0$ on the Helmholtz equation,
\be
(\nabla^2+k^2)\psi=0.
\label{helmholtz} 
\ee 
The wave function, $\psi$, and the energy eigenvalues, $E^{sq}_{m,n}$, are available in any standard text book of Quantum Mechanics and given by
\bes
\bea
\psi^{sq} &=& \frac{2}{L} \sin \frac{m \pi x}{L} \sin \frac{ n \pi y}{L}, \\
E^{sq}_{m,n} &=& \L \left( m^2 + n^2 \right)\qquad m,n=1,2,3, \cdots,
\label{squarespec}
\eea
\ees
where $\L = \frac{h^2}{8\mu L^2}$. The ground state energy is $E_{1,1} =2\L$, followed by the energy eigenvalues $E_{1,2} =5\L, E_{2,2} = 8\L, E_{1,3} = 10\L, E_{2,3} =13\L$ and so on. Without any loss of generality, we may normalize $L=1$. We immediately observe that the energy spectrum can be written in terms of $\mc{A}(m,n)$ as
\be 
E^{sq}_{m,n} = {\L}\mc{A} (m,n),
\ee
that is, the energy spectrum is in exact one-to-one correspondence with the amplitude-square of the corresponding periodic orbit in the classical 
billiard. Invariance of $\mc{A}(m,n)$ under the interchange $(m,n) \leftrightarrow (n,m)$ translates to the fact that the energy levels are doubly degenerate except for the cases $n = m$. On the other hand, invariance of $\mc{A}(m,n)$ under the transformation $x_0 \to x_0 + \D x, y_0 \to y_0 + \D y$ is interpreted on the quantum side as a phase transformation, $\psi \to \psi e^{i\t}$ that does not interfere with the energy levels\cite{Hsieh}. 

Having discussed the case of the square billiard with considerable detail, the rectangular one can be handled in a similar fashion and we omit its discussion. Instead, we turn to the case of a triangular billiard. As discussed earlier, there exist three integrable triangular billiards defined by the angles: $(\frac{\pi}{2}, \frac{\pi}{4}, \frac{\pi}{4}), (\frac{\pi}{3}, \frac{\pi}{3}, \frac{\pi}{3})$ and $(\frac{\pi}{2}, \frac{\pi}{3}, \frac{\pi}{6})$. Since triangular billiards are not as well-known as their square counterpart, they warrant a detailed discussion which we do in the following subsections. 

\begin{figure} [ht] 
\begin{center} 
\includegraphics[scale=0.45, angle=0]{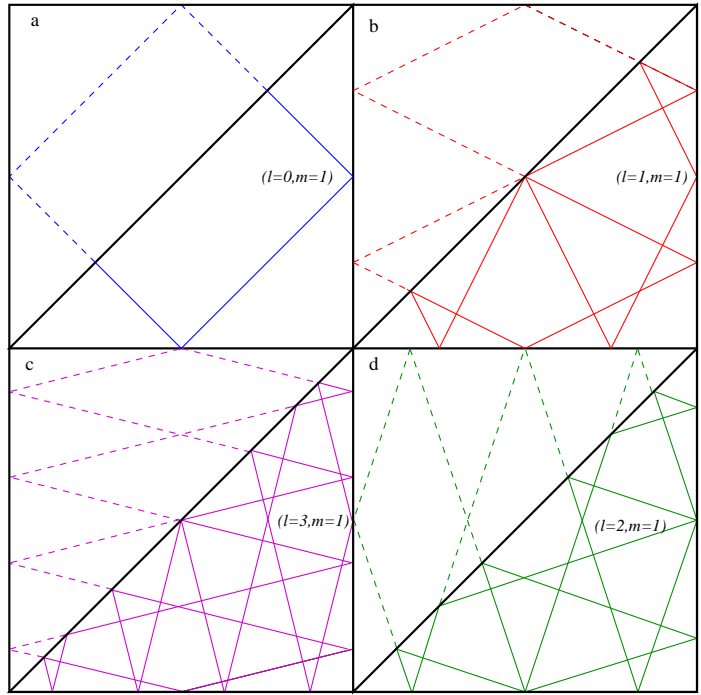} 
\caption{Periodic trajectories in a right-isosceles triangle billiard for different values of $(l,m)$.}
\label{isotriang}
\end{center}
\end{figure}
\subsection{ The right-isosceles triangle billiard}

The vertex angles of a right-isosceles triangle billiard are $(\frac{\pi}{2}, \frac{\pi}{4}, \frac{\pi}{4})$. Some sample periodic trajectories of a particle moving in this billiard are shown in Fig.\ref{isotriang}. Interestingly, all of them are obtained merely by folding the corresponding periodic trajectories of a square billiard. However, note that folding a square billiard trajectory with $m = n$, yields a trajectory in the triangular billiard that belongs to the Neumann spectrum and hence, we discard such trajectories for further discussion. The distinguishing feature of such a trajectory is that it will hit one of the boundaries normally and retrace the path back, an example of which is shown in Fig.\ref{isotriang}a. 
\begin{figure} [ht] 
\begin{center} 
\includegraphics[scale=0.7, angle=0]{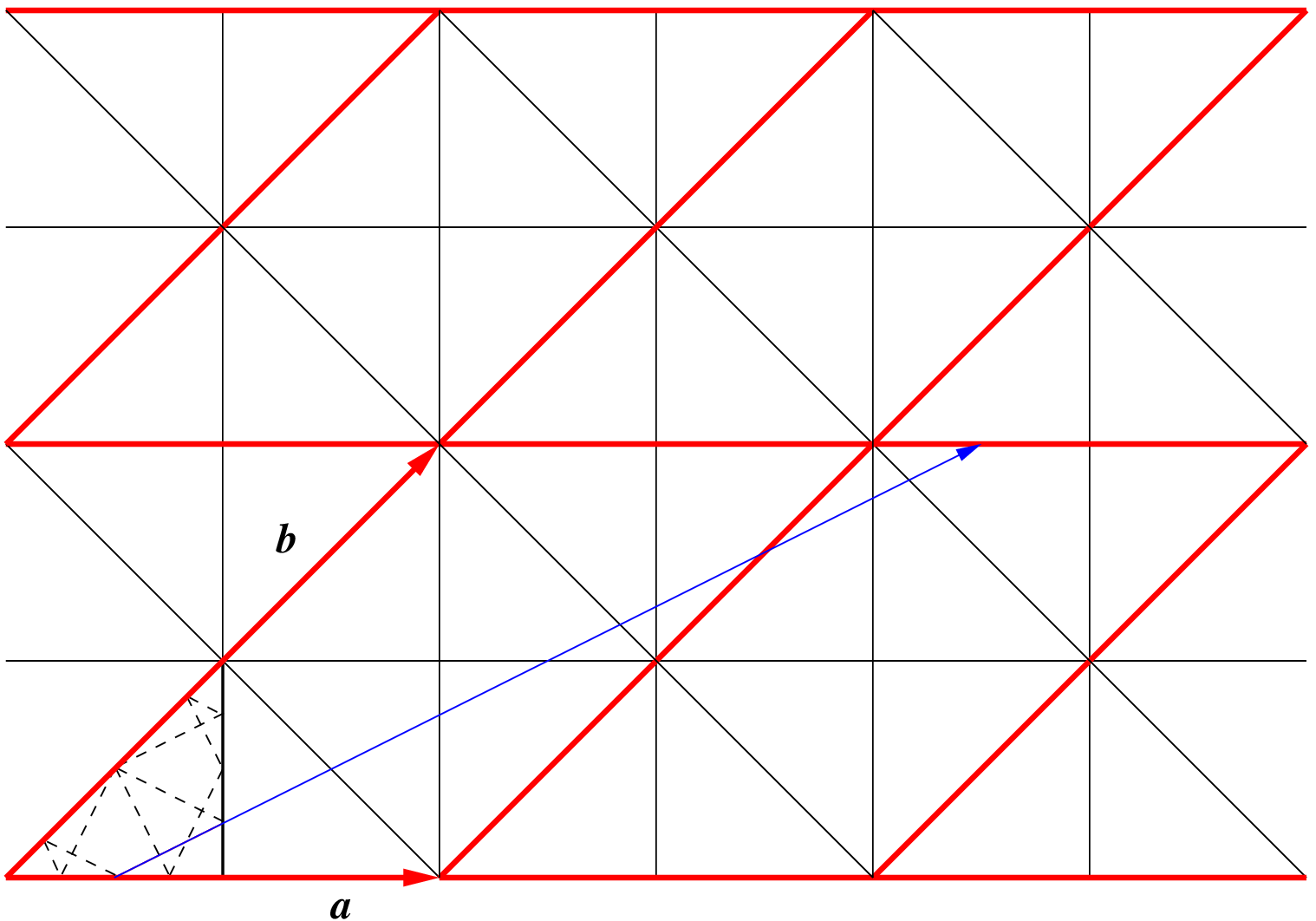} 
\caption{Construction of periodic orbit in a right-isosceles triangle billiard. The trajectory shown has $(l,m) = (1,1)$ and is the same as the one shown in Fig.\ref{isotriang}b.}
\label{isotriangor}
\end{center}
\end{figure}

Geometrically, we generate the trajectories using a square lattice (shown in red in Fig.\ref{isotriangor}) with basis vectors ${\bf {\it a}}=2L{\bf i}$ and ${\bf{\it b}}=2L({\bf i}+{\bf j})$. As in the case of the square billiard, here too, we immerse another self-similar lattice scaled by a factor half (shown in black) in it. Any vector (shown in blue), of the type ${\bf V}=l{\bf {\it a}}+m{\bf{\it b}}$, where $l$ and $m$ are two non-negative integers, in the parent lattice, can be folded as a periodic orbit in the fundamental region (shown in the left-bottom corner) of the daughter lattice, as shown in Fig.\ref{isotriangor}, by a dashed line. We denote the trajectory amplitude, $A(l,m)$ by the half length of the vector ${\bf V}$, i.e., $A(l,m)=\frac{1}{2}|{\bf V}|$ and 
\be
{\mc A}(l,m)=\left[A(l,m)\right]^2=(l^2+2m^2+2lm)L^2 \qquad {l,m=1,2,3, \cdots}.
\label{isorightamp}
\ee
Without any loss of generality, $L$ can be set to unity. Integers $l$ and $m$ are related to the number of collisions the particle
makes with the $x$-axis (horizontal boundary) and the hypotenuse of the billiard, which are given by $s$ and $h$, respectively. The relations are
\be
s=l+2m\qquad {\rm and}\qquad h=(l+1)(m+1)
\ee
which can be inverted to yield
\be
l=\frac{(s+1)-\sqrt{(s+3)^2-8h}}{2}\qquad{\rm and}\qquad 
m=\frac{(s-1)+\sqrt{(s+3)^2-8h}}{4} .
\ee
On the quantum side, we have to figure out the energy spectrum of a particle moving in a region having the shape of a right-isosceles triangle and having infinitely high potential wall along the boundary. The problem is also equivalent to finding the vibration modes of a uniform, flexible membrane, stretched under uniform tension and clamped on the boundary that has a shape of a right-isosceles triangle. The modes are obtained by separating out the temporal part and solving the Helmholtz equation with Dirichlet boundary condition \cite{Morse}. Upon comparing the problem with its square domain counterpart, we observe that the new boundary for this triangle is defined by the hypotenuse: $x=y$. This restricts the domain of the problem and the region $y> x$ is now inaccessible to the particle. Consequently, the symmetry $x \leftrightarrow y$ is spoiled and one loses the $m=n$ states from the spectrum of the square domain (\ref{squarespec}). Hence, the spectrum in this case is given by 
\be
E_{m,n}^{iso} = \L \left( m^2 + n^2 \right)\qquad m=1,2,3, \cdots;~n>m.
\label{isospecres}
\ee
We can alternatively impose the $n>m$ condition by writing $n=m+l$, where $l=1,2,3 \cdots$, so that (\ref{isospecres}) can be written in terms of two quantum numbers free from any constraint as follows
\be
E_{m,m+l}=E^{iso}_{l,m}=\L\left[m^2+(m+l)^2\right]=\L\left(l^2+2m^2+2lm\right)=\L{\mc A}(l,m);\qquad
{l,m=1,2,3, \cdots}.
\label{isospec}
\ee
 
Again, we have the surprising relation: the squares of the amplitudes, (\ref{isorightamp}), are in one-to-one correspondence with the quantum energy eigenvalues. The $m=n$ case of the square billiard would correspond to the $l=0$ case in the spectrum (\ref{isorightamp}) of the right-isosceles triangle, as shown Fig.\ref{isotriang}a. As mentioned earlier, these states would not fall in Dirichlet spectrum of the problem but would be included in the Neumann spectrum with identical boundary.
\begin{figure} [ht] 
\begin{center} 
\includegraphics[scale=0.45, angle=0]{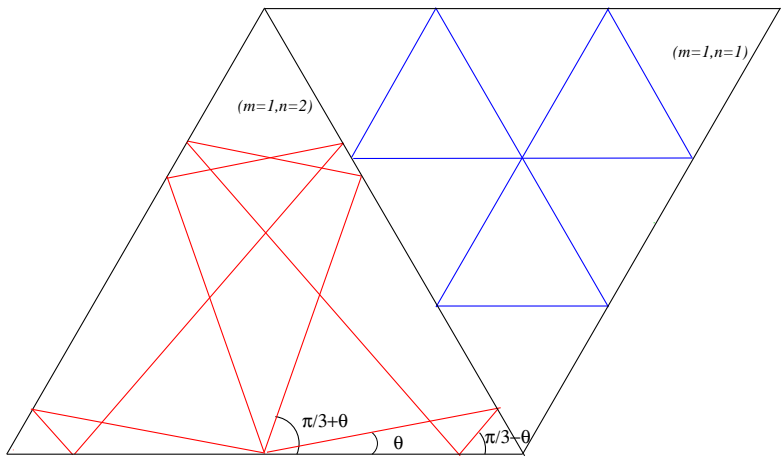} 
\caption{Periodic trajectories in equilateral triangle billiard.}
\label{equiorsm}
\end{center}
\end{figure}
\subsection{The equilateral triangle billiard}
An equilateral triangle billiard is also well-studied in both its classical and quantum incarnations \cite{Mathews, Pinsky, Krishna, McCartinA, Baxter, McCartinB}. The eigenspectrum for a quantum particle of mass $\mu$ trapped in an equilateral triangular box of side length $L$ was obtained in \cite{Mathews, Pinsky, Krishna}. The literature usually makes use of more than two quantum numbers and puts a constraint on them or uses two quantum numbers where one of them is dependent on the other in some way. Here we recast the results in terms of only two quantum numbers to write the complete eigenspectrum of the equilateral triangle problem without any restrictions on them, like the earlier two cases. The required bilinear form is given by \cite{McCartinB}
\be
E^{eq}_{m,n}=\L'(m^2+n^2+mn);\qquad
{m,n=1,2,3, \cdots},\qquad \L'\propto \frac{\pi^2\hbar^2}{\mu L^2}.
\label{equispec}
\ee
\begin{figure} [ht] 
\begin{center} 
\includegraphics[scale=0.6, angle=0]{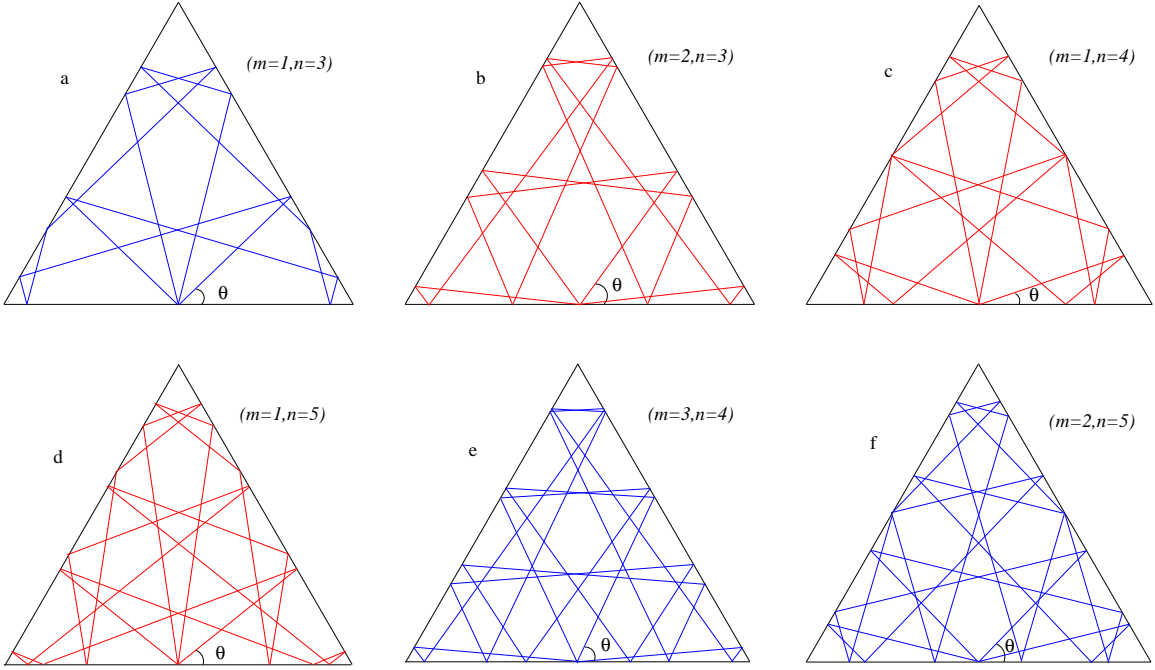} 
\caption{ Periodic trajectories in an equilateral triangle billiard. Length of each side is normalized to $2/\sqrt{3}$. The $m$ and $n$ values for the trajectories are shown in the figure ($m<n$). The amplitudes, tangents of the shooting angles and the number of collisions are respectively for: a) $\sqrt{13},~\frac{5}{3\sqrt{3}},~14$;~ b) $\sqrt{19},~\frac{7}{3\sqrt{3}},~16$; ~ c) $\sqrt{21},~\frac{3}{5\sqrt{3}},~18$; d) $\sqrt{31},~\frac{7}{5\sqrt{3}},~22$; e) $\sqrt{37},~\frac{11}{3\sqrt{3}},~22$;~ f) $\sqrt{39},~\frac{9}{5\sqrt{3}},~24$;}
\label{equior}
\end{center}
\end{figure}
For the classical equilateral triangle billiard, the periodic orbits are classified in \cite{Baxter}. In the following, we systematically construct these orbits in a unified approach using our method of geometric construction. Fig.\ref{equiorsm} shows the shortest length orbit and the next one. For convenience, we normalize the length of each side to $2/\sqrt{3}$, so that the shortest periodic orbit, shown in blue, has amplitude $A = \sqrt{3}$. The next trajectory (shown in red) has amplitude $A =\sqrt{7}$. The shooting angle for the shortest periodic trajectory is $\pi/3=\tin (\sqrt{3})$ with respect to horizontal, whereas, for the next orbit shown in Fig.\ref{equiorsm}, $\theta=\tin (\frac{1}{3\sqrt{3}})$ . While it is only natural that the same shooting angle will preserve the orbit length, in this particular case, due to the shape of the billiard and the laws of reflection, there are, in fact, three different shooting angles which generate closed orbits of identical length. Fig.\ref{equior} shows some of the periodic orbits beyond the shortest and the next-to-shortest ones - the corresponding shooting angles are mentioned in the figure caption.

\begin{figure} [ht] 
\begin{center} 
\includegraphics[scale=0.45, angle=0]{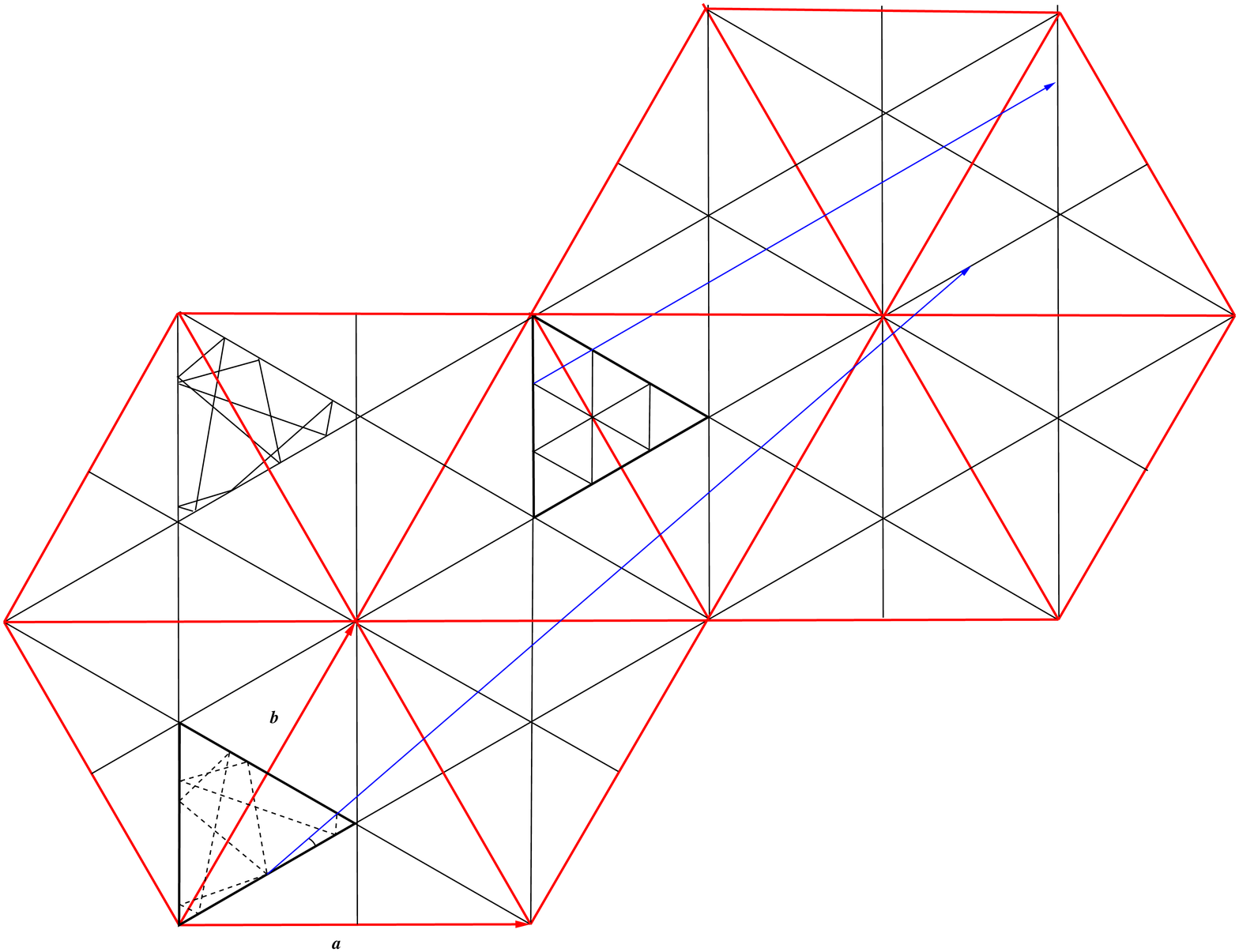} 
\caption{Construction of periodic orbits in an equilateral triangle billiard. The shorter trajectory (continuous lines) has $m=1$ and $n=1$ while the longer one (dashed lines in the bottom) has $m=1$ and $n=2$. The third trajectory on the top is equivalent to the $(m=1,n=2)$ case and is just shot from a different initial position.} 
\label{tranlat}
\end{center}
\end{figure}
\begin{figure} [ht] 
\begin{center} 
\includegraphics[scale=0.45, angle=0]{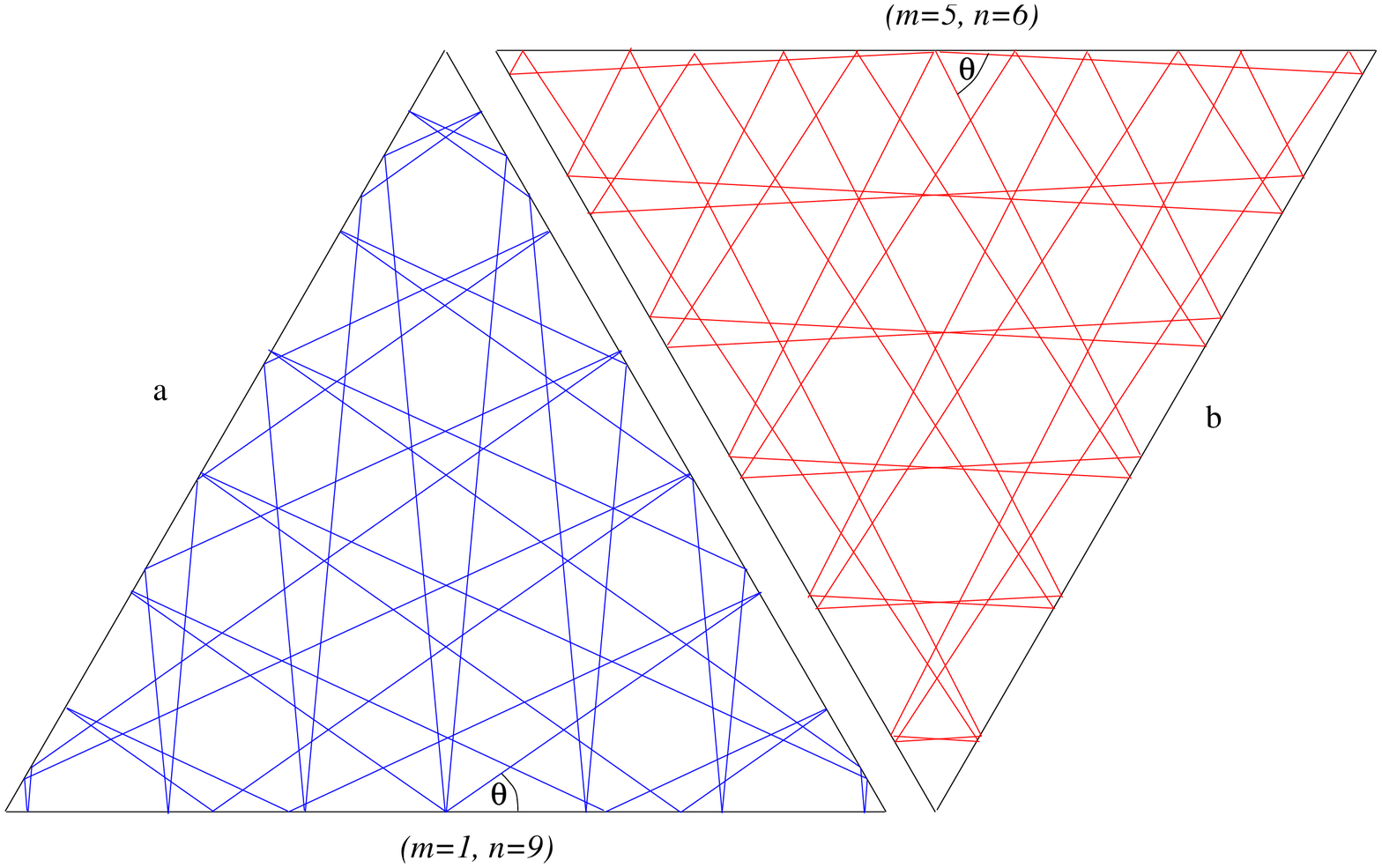} 
\caption{Two degenerate trajectroies with amplitude $\sqrt{91}$ 
corresponding to $(m,n)$ values $(1,9)$ and $(5,6)$. Tangent of the shooting angle and number of collisions are respectively for a) $\frac{11}{9\sqrt{3}},~38$; b) $\frac{17}{5\sqrt{3}},~34$.} 
\label{degenamp}
\end{center}
\end{figure}

Geometrically, a periodic orbit is obtained by drawing a vector, ${\bf V}$, (shown in blue in Fig.\ref{tranlat}), between two equivalent points on a parent lattice (shown in red) and folding the vector to the fundamental region of the daughter lattice. In this case, the parent lattice has lattice vectors ${\bf {\it a}}=2L{\bf i}$ and ${\bf {\it b}}=L({\bf i}+{\sqrt{3}}{\bf j})$. This makes the sides of the fundamental region of the daughter lattice (shown in black) equal to $2L/\sqrt{3}$. Each orbit in the daughter lattice corresponds to a vector in the parent lattice of the form ${\bf V}=m{\bf a}+n{\bf b}$, where $m$ and $n$ are two positive integers. Thus, like in the earlier cases, each orbit is characterized by a pair of integers $(m,n)$. It is at once evident that the orbits $(m,n)$ and $(n,m)$ will be degenerate owing to the reflection symmetry of the geometry. In addition to this expected degeneracy, the spectrum of the amplitudes also enjoys accidental degeneracies, where two completely different trajectories suffering different number of collisions have the same amplitude. The first such occurrence comes for the combinations, $(5,6)$ and $(1,9)$, which suffer $34$ and $38$ collisions respectively. The trajectories are shown in Fig.\ref{degenamp}. There can also be pairs of trajectories where each member of a pair has different amplitude, but both members suffer the same number of collisions in one closed loop. This type of degeneracy is shown in Fig.\ref{equior}d and Fig.\ref{equior}e, where the number of collisions for both the trajectories is $22$. 
We denote trajectory amplitude, $A(m,n)$ by the half length of the vector ${\bf V}$, i.e., $A=\frac{1}{2}|{\bf V}|$ and 
\be
{\mc A}(m,n)=\left[A(m,n)\right]^2=(m^2+n^2+mn)L^2 \qquad {m,n=1,2,3, \cdots}.
\label{equirightamp}
\ee
 
Setting $L$ equal to unity, we again find that there is one-to-one correspondence between the energy spectrum of a quantum particle trapped in an equilateral triangle and squared amplitudes of periodic orbits in a classical billiard of the same shape:
\be
E^{eq}_{m,n}=\L'(m^2+n^2+mn)=\L'{\mc A}(m,n);\qquad
{m,n=1,2,3, \cdots}.
\label{eqispec}
\ee

Table \ref{eqlist} shows the details of periodic trajectories in an equilateral triangle billiard. We have considered only set of co-primes, $(m,n)$ with $ m\le n $. The spectrum would also contain combinations of the form $(pm,pn)$, where $p$ is a positive integer. If $\theta$ is the shooting angle measured from the $x$-axis (with which one of the sides of the equilateral triangle is aligned), then the shooting angles $\alpha=(\pi/3)-\theta$ and $\beta=(\pi/3)+\theta$ produce the same orbit. Angles $\pi-\theta$, $\pi-\alpha$ and $\pi-\beta$ are also members of the same orbit. We present $|\tan\theta|,~|\tan\alpha|$ and $|\tan\beta|$ values for each combination. The shooting angles are always of the form, $\tin\left[{p}/{(q{\sqrt 3})}\right]$, where the combinations $(p,q)$ are obtained from $(m,n)$ in the following way for the three angles:
\be
|\tan\theta|=\frac{m+2n}{{\sqrt 3}n},\qquad |\tan\alpha|= \frac{2m+n}{{\sqrt 3}m}, \qquad |\tan\beta|= \frac{n-m}{{\sqrt 3}(m+n)}. \ee
 
\begin{table}
\caption{The combinations $(m,n)$, amplitude-square, shooting angles for periodic orbits in equilateral triangle billiard.} 
\begin{center}
\begin{tabular}[ht]{*{5}{c}}
\toprule
\midrule
Combinations & Amplitude-square & Degen- & No. of Collisions & Combinations $(p,q)$ for \\
$(m,n)(m\le n)$ & $=m^2+mn+n^2$ & eracy& $=2(m+2n)$ & 
$|\tan\theta|,~|\tan\alpha|,~
|\tan\beta|$\\ \toprule
(1,1) & 3 & 1 & 6 & (3,1), (3,1), (0,2) \\ \midrule
 (1,2) & 7 & 2 & 10& (5,1), (2,1), (1,3)\\ \midrule 
 (1,3) & 13 & 2 & 14 & (7,1), (5,3), (1,2)\\ \midrule
(2,3) & 19 & 2 & 16 & (4,1), (7,3), (1,5)\\ \midrule
(1,4) & 21 & 2 & 18 & (9,1), (3,2), (3,5)\\ \midrule
(1,5) & 31 & 2 & 22 & (11,1), (7,5), (2,3)\\ \midrule
(3,4) & 37 & 2 & 22 & (11,3), (5,2), (1,7) \\ \midrule
(2,5) & 39 & 2 & 24 & (6,1), (9,5), (3,7)\\ \midrule
 (1,6) & 43 & 2 & 26 & (13,1), (4,3), (5,7)\\ \midrule
(3,5) & 49 & 2 & 26 & (13,3), (11,5), (1,4) \\ \midrule
 (1,7) & 57 & 2 & 30 & (15,1), (9,7), (3,4)\\ \midrule
(4,5) & 61 & 2 & 28 & (7,2), (13,5), (1,9) \\ \midrule
(2,7) & 67 & 2 & 32 & (8,1), (11,2), (5,9)\\ \midrule
(1,8) & 73 & 2 & 34 & (17,1), (5,4), (7,9)\\ \midrule
 (3,7) & 79 & 2 & 34 & (17,3), (13,7), (2,5)\\ \midrule
 (5,6) & 91 & 2 & 34 & (17,5), (8,3), (1,11)\\ 
(1,9) & 91 & 2 & 38 & (19,1), (11,9), (4,5) \\ \midrule
(4,7) & 93 & 2 & 36 & (9,2), (15,7), (3,11) \\ \midrule
 (3,8) & 97 & 2 & 38 & (19,3), (7,4), (5,11)\\ \midrule
(2,9) & 103 & 2 & 40 & (10,1), (13,9), (7,11)\\ \midrule
(5,7) & 109 & 2 & 38 & (19,5), (17,7), (1,6) \\ \midrule
 (1,10) & 111 & 2 & 42 & (21,1), (6,5), (9,11)\\ \midrule
\bottomrule
\end{tabular}
\end{center}
\label{eqlist}
\end{table}

\subsection{The hemi-equilateral triangle billiard}
The last integrable polygonal billiard has the shape of a hemi-equilateral triangle. The angles of this triangular billiard are $(\pi/6,\pi/3,\pi/2)$.
The energy eigenspectrum of this billiard is a subset of the spectrum of an equilateral triangle. If we consider an equilateral triangle with the base coinciding with the $x$-axis and the mid point of the base as origin, the spectrum of the hemi-equilateral triangle is obtained by filtering out the odd parity eigenfunctions, i.e., the ones for which $\psi(-x)=-\psi(x)$. In other words, we impose another Dirichlet condition on the $y$-axis which serves as the new boundary of the hemi-equilateral triangle. This is achieved by the removal of $m=n$ states in (\ref{equispec}). So we have the spectrum
\be
E^{hemi}_{m,n} = \L' \left( m^2 + n^2+mn \right)\qquad m=1,2,3, \cdots;~n>m.
\label{hemispecres}
\ee
We can extract out the $n>m$ states we are seeking by writing $n=m+l$, where $l=1,2,3 \cdots$, so that (\ref{hemispecres}) can be written without any constraints on the two quantum numbers as follows
\be
E^{eq}_{m,l+m}=E^{hemi}_{l,m}=\L'\left[m^2+(m+l)^2+m(m+l)\right] = \L'(l^2+3m^2+3lm); \qquad
{l,m=1,2,3, \cdots}.
\label{hemispec}
\ee
\begin{figure} [ht] 
\begin{center} 
\includegraphics[scale=0.71, angle=0]{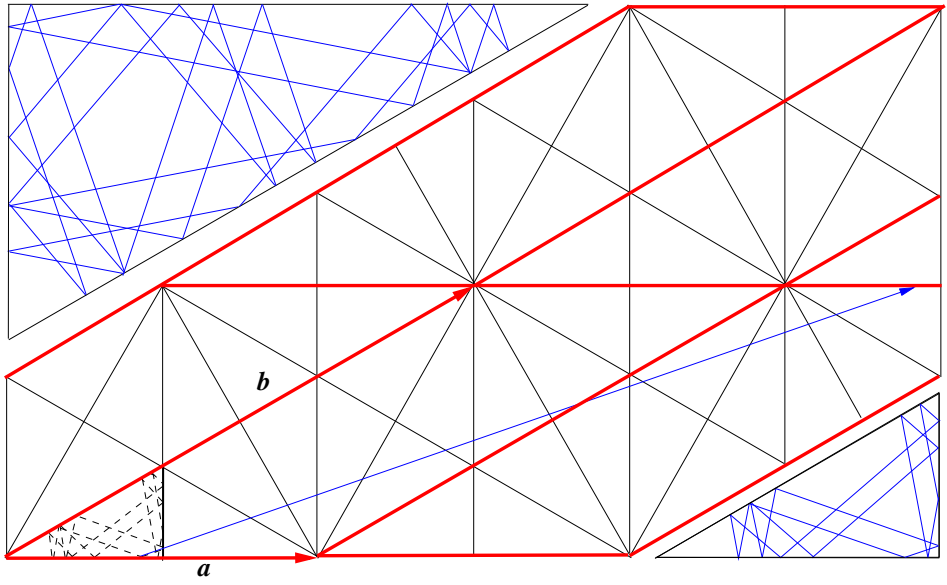} 
\caption{Construction of periodic orbits in a hemi-equilateral triangle billiard. The details of the folded, original trajectory with $A=\sqrt {7}$ and $l=1$ and $m=1$ is shown in the right lower corner. Another orbit with $l=3$ and $m=1$, and $A = \sqrt{21}$, is shown in the upper left corner.} 
\label{hemiequi}
\end{center}
\end{figure}
Let us now turn to the classical side of the problem. On the classical side, periodic orbits are obtained by folding the $(m,n)$ orbits of an equilateral triangle with $m \neq n$, just like the spectrum of a right-isosceles triangle was obtained from the square billiard by filtering out the $m = n$ states. For the geometric construction, we first create the parent lattice, as shown in red in Fig.\ref{hemiequi} and in this we embed a hemi-equilateral triangular lattice, shown in black, and which is, of course, reflective. The basis vectors for the parent lattice are given by ${\bf {\it a}}=2L{\bf i}$ and ${\bf {\it b}}=\sqrt{3}L(\sqrt{3}{\bf i}+{\bf j})$. Any vector of the type, ${\bf V}=l{\bf {\it a}}+m{\bf {\it b}}$ can be folded into the fundamental region of the daughter lattice, as shown in Fig.\ref{hemiequi}. The amplitude of the orbit is $A(l,m)$ defined, as usual, as $A=\frac{1}{2}|{\bf V}|$ and
\be
{\mc A}(l,m)=\left[A(l,m)\right]^2=(l^2+3m^2+3lm)L^2 \qquad {l,m=1,2,3, \cdots}.
\label{hemiequiamp}
\ee
The number of collisions in this case is given by $2(3l+5m)$. Setting $L$ equal to unity, we again find that there is one-to-one correspondence between the energy spectrum of a quantum particle trapped in a hemi-equilateral triangle and squared amplitudes of periodic orbits in identical classical billiard:
\be
E^{hemi}_{l,m}=\L'(l^2+3m^2+3lm)=\L'{\mc A}(l,m);\qquad
{l,m=1,2,3, \cdots}.
\label{hemieqispec}
\ee
We summarize the results of this section compactly in table \ref{2dsumm} with $L=1$. For convenience, we have relabeled some of the expressions so that all the results now contain two positive integers $l$ and $m$. For all the cases, $2A = |\mb{V}|$, where ${\bf V}=l{\bf {\it a}}+m{\bf {\it b}}$. The integers $l$ and $m$ take values $1,2,3, \cdots$. The Neumann spectra for the integrable boundaries are also given by the same bilinears given in table \ref{2dsumm} with the only difference that in this case either of $l$ and $m$ can assume
the value 0. This shows that the Dirichlet spectrum is a subset of the Neumann spectrum for all the cases.

\bigskip
\begin{table}
\caption{Lattice vectors, number of collisions and the amplitude-squares
 for 2D integrable billiards.}
\begin{center}
\begin{tabular}[ht]{*{5}{c}}\toprule
\midrule
Shape & Lattice vectors & Number of
collisions &
Amplitude$^2$=$\frac{|{\bf V}|^2}{4}$\\
\midrule
Square & ${\bf {\it a}}=2{\bf i}$; ${\bf{\it b}}=2{\bf j}$ & $2(l+m)$ & $l^2+m^2$\\ \midrule
Right-isosceles triangle & ${\bf {\it a}}=2{\bf i}$; 
${\bf{\it b}}=2({\bf i}+{\bf j})$ & $3l+5m+lm+1$ 
& $l^2+2lm+2m^2$\\ \midrule
Equilateral triangle & ${\bf {\it a}}=2{\bf i}$; ${\bf {\it b}}={\bf i}+\sqrt{3}{\bf j}$ & $2(l+2m),~l\leq m$ &$l^2+lm+m^2$ \\ \midrule
Hemi-equilateral triangle & ${\bf {\it a}}=2{\bf i}$; ${\bf {\it b}}=\sqrt{3}(\sqrt{3}{\bf i}+{\bf
 j})$ & $2(3l+5m)$ & $l^2+3lm+3m^2$ \\ \midrule
\bottomrule
\end{tabular}
\end{center}
\label{2dsumm}
\end{table}

\begin{figure} [ht] 
\begin{center} 
\includegraphics[scale=0.60, angle=0]{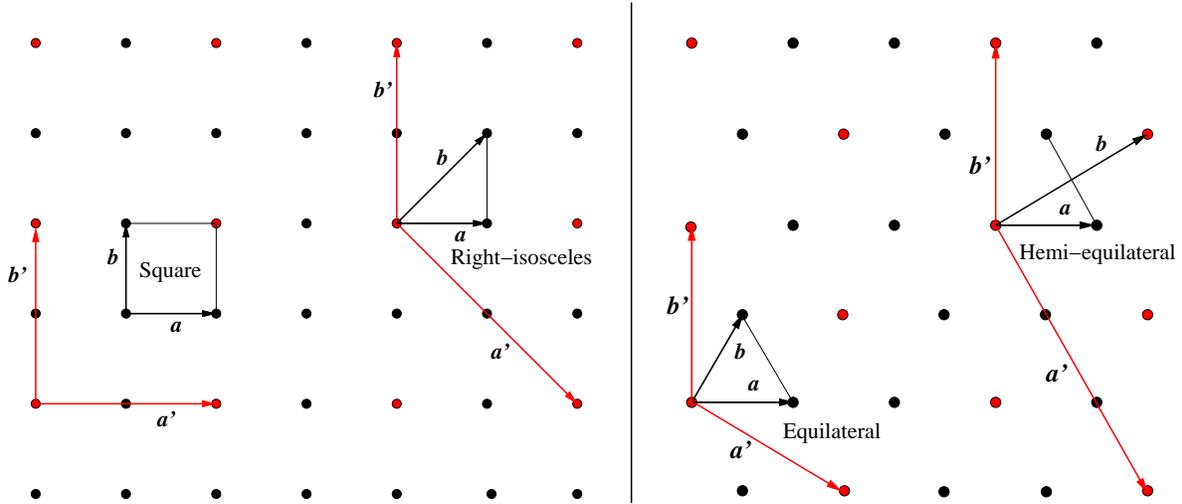} 
\caption{Left: Square lattice; Right: Hexagonal lattice. The daughter lattice and the fundamental zones are shown in black; Red points are accessible to the reciprocal (or the parent) lattice for which the lattice vectors are shown with red arrows. Reciprocal lattice constant is normalized to ${\sqrt 2}$.}
\label{lattice}
\end{center}
\end{figure}

This concludes our discussion of the four integrable geometries for which periodic orbits exist. Fig.\ref{lattice} shows all four sets of lattice
vectors for $2$-dimensional integrable domains. 
It is interesting to note that the two cases - the square and the right-isosceles triangle are generated from a square lattice primitive cell basis vectors. The remaining two, i.e., the equilateral triangle and the hemi-equilateral triangle are generated from a hexagonal lattice primitive cell basis vectors. The body-centered square lattice again produces a square lattice of reduced lattice constant (reduced by a factor of ${\sqrt 2}$), so it does not generate a new domain. The rectangular domain, with primitive cell basis vectors as lattice generators, is again exactly solvable - it is a trivial extension of the square domain. This exhausts all the possibilities for two dimensions. For every case, the daughter lattice (shown in black with lattice vectors ${\it{\bf a}}$ and ${\it {\bf b}}$) can be obtained from the parent lattice (shown in red with lattice vectors ${\it {\bf a'}}$ and ${\it {\bf b'}}$). In all the cases, the parent lattice is the reciprocal lattice. We normalize the reciprocal lattice constant to $\sqrt{2}$, i.e., $|{\it {\bf b'}}|=\sqrt{2}$. The moduli, $|{\it {\bf V'}}|$, of the vectors formed by reciprocal lattice vectors, ${\it {\bf V'}}=l{\it {\bf a'}}$+m${\it {\bf b'}}$ are isospectral with the periodic trajectory amplitudes $|\mb{V}|$ if the
parameter $l$ is constrained such that $l>m$ except for the square case where no such restriction is required. This result is summarized in table \ref{recdir}. 

\bigskip
\begin{table}
\caption{Reciprocal lattice vectors, their square-moduli 
 for 2D integrable billiards.}
\begin{center}
\begin{tabular}[ht]{*{5}{c}}\toprule
\midrule
Shape & Reciprocal lattice vectors & $\frac{|{\it{\bf V'}}|^2}{2}$,\scriptsize{ ${\it {\bf V'}}=l{\it {\bf a'}}$+m${\it {\bf b'}}$} &
Amplitude$^2$=$\frac{|{\bf V}|^2}{4}$\\
\midrule
Square & ${\bf {\it a'}}=\sqrt{2}{\bf i}$; ${\bf{\it b'}}=\sqrt{2}{\bf j}$ & $l^2+m^2$ & $l^2+m^2$\\ \midrule
Right-isosceles triangle & ${\bf {\it a'}}=\sqrt{2}({\bf i}-{\bf j})$; 
${\bf{\it b'}}=\sqrt{2}{\bf j}$ & $2l^2-2ml+m^2$ 
& $l^2+2lm+2m^2$\\ \midrule
Equilateral triangle & ${\bf {\it a'}}=\frac{1}{\sqrt{2}}(\sqrt{3}{\bf i}-{\bf j})$; ${\bf {\it b'}}=\sqrt{2}{\bf j}$ & $l^2-lm+m^2$ &$l^2+lm+m^2$ \\ \midrule
Hemi-equilateral triangle & ${\bf {\it a'}}=\sqrt{\frac{3}{2}}({\bf
 i}-\sqrt{3}{\bf j})$; ${\bf {\it b'}}=\sqrt{2}{\bf j}$ & $3l^2-3ml+m^2$ & $l^2+3lm+3m^2$ \\ \midrule
\bottomrule
\end{tabular}
\end{center}
\label{recdir}
\end{table}

\section{ Billiards in three dimensions}
In this section, we extend our method of geometric construction to $3$-dimensional billiards. In three dimensions, only the cube, among the convex regular polyhedra (Platonic solids), can fill the $3$-dimensional space without leaving holes. The problem of space-filling in three dimension with congruent copies dates back to Hilbert who, more than a hundred year ago, in his eighteenth problem, addressed related issues. Restricting to tetrahedral tiling, Sommerville found four types \cite{Sommerville1,Sommerville2} and believed the list was complete. Davies\cite{Davies} and Baumgartner\cite{Baumgartner1,Baumgartner2} discovered three of these tetrahedra independently. Further, Baumgartner found a new one which was not present in Sommmervile's list, thereby, extending the list to five space-filling tetrahedra in the literature. Goldberg showed that three of these are special cases of three new infinite one-parameter family of space-filling tetrahedra \cite{Goldberg1, Goldberg2}. Among these space-filling tetrahedra, only three could fill the space with reflective tesselations\cite{Turner}.
\begin{figure} [ht] 
\begin{center} 
\includegraphics[scale=0.75, angle=0]{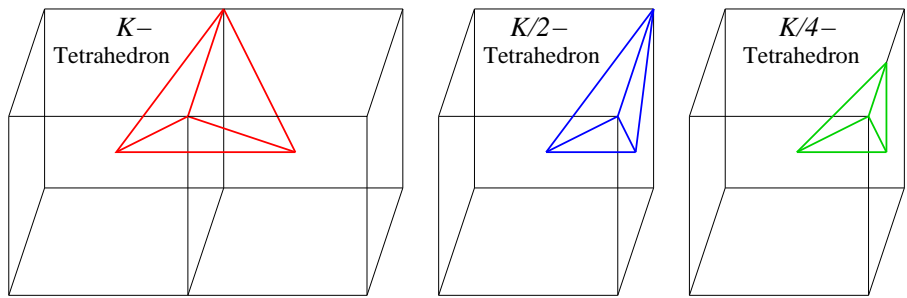} 
\caption{The three tetrahedra which fills the $3$-dimensional space without gaps with reflective tesselations. Each cube has a side length of 2 units.}
\label{somertetra}
\end{center}
\end{figure}
The first of these tetrahedra has two opposite edges with length $2$ units and the dihedral angles for these edges are the same and equal to $\pi/2$. The other four edges are of the same length equal to ${\sqrt 3}$ and the dihedral angles at these edges are equal to $\pi/3$.
It was shown by Krishnamurthy et. al. that the Helmholtz equation in such a domain with Dirichlet boundary condition is an exactly solvable problem and the resulting spectrum was reported in \cite{Krishna}. Hereafter, following \cite{Turner}, we refer to such a tetrahedron as $K$-tetrahedron. The other two space filling tetrahedra are just half and one fourth of a $K$-tetrahedron and we refer to them as $K/2$- and $K/4$-tetrahedra respectively. The three tetrahedra are shown inside unit cubes in Fig.\ref{somertetra}. The $K/2$-tetrahedron has edge length ratios equal to (2, ${\sqrt 3}$, ${\sqrt 3}$, ${\sqrt 2}$, ${\sqrt 2}$, 1). There are three right angle dihedrals which are at 
concurrent edges and the rest three dihedral angles are $\pi/3$, $\pi/3$ and $\pi/4$. The $K/4$-tetrahedron has edge length rations equal to (${\sqrt 3}$, ${\sqrt 2}$, ${\sqrt 2}$, 1, 1, 1). There are again three right angle dihedrals but in this case they are not at concurrent edges and the rest three dihedral angles are $\pi/3$, $\pi/4$ and $\pi/4$. For the construction of periodic orbits in these polyhedra, we proceed exactly like we did for the $2$-dimensional cases. We conjecture that the energy eigenspectrum of a particle trapped in such a polyhedron is in one-to-one correspondence with the amplitude-square spectrum of the periodic orbits. We then verify our conjecture with analytic and numerical results.

\subsection{The cubic billiard} 
The generalization from square to cube is pretty straight forward. Helmholtz equation, (\ref{helmholtz}), for a cubic domain with Dirichlet 
boundary condition is exactly solvable by the separation of variables. The wave functions and the corresponding energy eigenvalues for the same are given by 
\bes
\bea
\psi^{cube} &=& \frac{2\sqrt{2}}{L\sqrt{L}} \sin \frac{l \pi x}{L} \sin \frac{ m \pi y}{L}
 \sin \frac{ n \pi z}{L} \\
E^{cube}_{l,m,n} &=& \L \left( l^2+ m^2 + n^2 \right);\qquad l,m,n=1,2,3, \cdots.
\label{cubespec}
\eea
\ees

\begin{figure} [ht] 
\begin{center} 
\includegraphics[scale=0.8, angle=0]{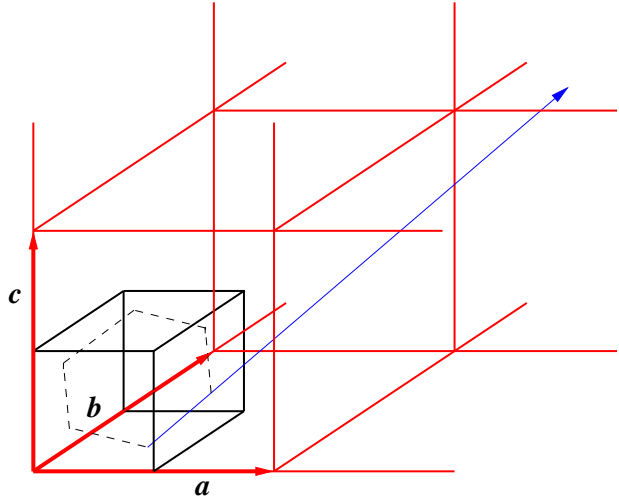} 
\caption{{\label{cubelatt}} Construction of a periodic orbit in a cubic billiard. Here $l=1$, $m=1$ and $n=1$ and the amplitude is ${\sqrt 3}$.} 
\end{center}
\end{figure}

On the classical side, the bouncing ball problem inside a cubic billiard has its traces in the mathematical notes of Lewis Carroll, where a solution of a cyclic path was sorted with the constraint that each segment of the path, between two successive reflections from the walls, has equal length \cite{Gardner}. In the general problem, the equal segment length constraint is removed, and one only looks for cyclic trajectories. The parent cubic lattice is easy to construct with lattice vectors given by, ${\bf {\it a}}=2L{\bf i}$, ${\bf{\it b}}=2L{\bf j}$ and ${\bf{\it c}}=2L{\bf k}$, where ${\bf i}$, ${\bf j}$ and ${\bf k}$ are unit vectors along the $x$, $y$ and $z$-axes respectively. The daughter lattice (which is also cubic) is immersed in the parent with lattice constant $L$, which is half of the parent lattice constant. Fig.\ref{cubelatt} shows the construction with the shortest trajectory folded in the fundamental region of the daughter lattice. Likewise, any vector, ${\bf V}$, of the form, $l{\bf {\it a}}+m{\bf{\it b}}+n{\bf{\it c}}$, can be folded as a periodic trajectory in the smaller lattice. 

\subsection{The $K$-tetrahedron billiard}
Helmholtz equation in a $K$-tetrahedron domain with Dirichlet boundary condition was exactly solved in \cite{Krishna}. The energy spectrum was given in a bilinear form involving three quantum numbers $l$, $p$ and $q$ with the condition $q > p > l; {l,p,q=1,2,3,\cdots}$,
\be 
E^K_{l,p,q}={\L}''\left[3(l^2+p^2+q^2)-2(lp+pq+lq)\right],
\label{ktetrasy}
\ee
where $\L''$ is a constant depending on Planck's constant, dimensions of the the domain and the mass of the particle. The expression for energy is symmetric under the interchange $l\leftrightarrow p$, $l\leftrightarrow q$ or $p\leftrightarrow q$. However, this symmetry is deceptive and not present in the problem. To realize this, we just need to recall the constraint $q > p > l$. The relabeling, $p=m+l$ and $q=n+p=l+m+n$ with $m,n=1,2,3\cdots$, takes care of the constraints on $p$ and $q$ and the expression for the energy becomes 
\be
E^K_{l,m,n}={\L}''[3l^2+4m^2+3n^2+4lm+2ln+4mn] \qquad l,m,n=1,2,\cdots.
\label{ktetra}
\ee 
Expression (\ref{ktetra}) is now symmetric under the interchange of $l$ and $n$. In the next section, we shall exploit this symmetry to obtain a new integrable $3$-dimensional domain. We conjecture that in $3$-dimensional integrable domains also, the energy eigenvalues are in exact one-to-one correspondence with amplitude-squares of the classical periodic orbits in the same domain. To establish this, we take recourse to our method of geometric construction. 

The geometric construction for the bilinear (\ref{ktetra}) is realized by the following primitive cell basis vectors of a body-centered cubic lattice, ${\bf{\it a}}=2({\bf i}+{\bf j}+{\bf k}); {\bf {\it b}}=4 {\bf j}$ and ${\bf {\it c}}=2(-{\bf i}+{\bf j}+{\bf k})$. The amplitude is given by $A = \h |\mb{V}|$, where ${\bf V}=l{\bf {\it a}}+m{\bf {\it b}}+n{\bf{\it c}}$. The integers $l$, $m$ and $n$ take values $1,2,3, \cdots$. The square of the amplitude is given by $|{\bf V}|^2$/4. It can be easily checked that $|{\bf V}|^2$ for this case is proportional to $E^K_{l,m,n}$, given in (\ref{ktetra}).

\subsection{The $K/2$-tetrahedron billiard}
We notice that the $K$-tetrahedron has two identical mirror symmetry planes, which are mutually orthogonal. Each of these passes through one of the longer edges (i.e., edges with length $2$ units) and the mid-point of the remaining longer edge. If one divides the $K$-tetrahedron through one of these planes, one obtains two $K/2$-tetrahedra. We exploit this symmetry to obtain the energy eigenspectra of a particle confined in a $K/2$-tetrahedra with Dirichlet boundary condition. This is achieved by the removal of degeneracies in the spectrum of $K$-tetrahedra. The method is identical to the one described for the $2$-dimensional cases, i.e., from square to right-isosceles triangle or, from equilateral triangle to hemi-equilateral triangle. We remove the $l\leftrightarrow n$ symmetry from the expression (\ref{ktetra}) by substituting
$l=n+p;~p=1,2,3\cdots$. This enforces the condition $l>n$ always. After substitution, we arrive at
\be
E^{K/2}_{{n+p},m,n}={\L}''\left[3(n+p)^2+4m^2+3n^2+4(n+p)m+2(n+p)n+4mn\right];\qquad p=1,2,3\cdots.
\label{k2tetra}
\ee 
Simplifying (\ref{k2tetra}) and relabeling $m\rightarrow l$, $n\rightarrow m$, $p\rightarrow n$, the exact energy eigenspectrum for $K/2$-tetrahedron is given by,
\be
E^{K/2}_{l,m,n}={\L}''\left[4l^2 + 8m^2 + 3n^2 + 8lm+4ln +8mn \right];\qquad l,m,n=1,2,3\cdots.
\label{k2ftetra}
\ee
The geometric construction for the bilinear (\ref{k2ftetra}) is realized by the following primitive basis vectors of a body-centered cubic lattice: ${\bf{\it a}}=4{\bf i}, {\bf {\it b}}=4({\bf i}+ {\bf j})$ and ${\bf {\it c}}=2({\bf i}+{\bf j}+{\bf k})$. Again, the amplitude is $\h |\mb{V}|$, with ${\bf V}=l{\bf {\it a}}+m{\bf {\it b}}+n{\bf{\it c}}$.

\subsection{The $K/4$-tetrahedron billiard}
In the previous section, we exploited one of the mirror symmetries of the $K$-tetrahedron to obtain a $K/2$-tetrahedron. The residual mirror plane is still available in the $K/2$-tetrahedron. In this section, we exploit this symmetry to obtain a $K/4$-tetrahedron, as shown in Fig.\ref{somertetra}. This is achieved by chopping the $K/2$-tetrahedron through the mirror plane into two identical halves. Each half so created is a $K/4$-tetrahedron. 
 
To obtain the energy eigenspectrum for this domain, we start with the $K/2$-spectrum, (\ref{k2ftetra}), and put $n=2p;,~p=1,2,3, \cdots$, that is allowing only even integers for the third direction. With this substitution, we obtain the exact expression of energy spectrum for a $K/4$-tetrahedron:
\be
E^{K/4}_{l,m,p}=4{\L}''\left[l^2+2m^2+3p^2+2lm+2lp+4mp\right];\qquad l,m,p=1,2,3, \cdots.
\label{k4tetra}
\ee 
Absorbing the overall factor of $4$ in the constant $\L''$ and relabeling $p\rightarrow n$, leads us to the following expression,
\be
E^{K/4}_{l,m,n}={\L}''\left[l^2+2m^2+3n^2+2lm+2ln+4mn\right];\qquad l,m,n=1,2,3, \cdots.
\label{k4ftetra}
\ee
Note that (\ref{k4ftetra}) can also be directly obtained from the cube spectrum (\ref{cubespec}) with the restriction $l>m>n$.

The geometric construction for this case readily follows from the $K/2$ basis vector construction. This is achieved by halving the
lattice vectors ${\bf {\it a}}=4{\bf i}$ and ${\bf {\it b}}=4({\bf i}+{\bf j})$ of $K/2$-construction (since the parameter $n$ corresponding to the basis vector ${\bf {\it c}}$ was doubled). The required basis vectors so obtained are ${\bf {\it a}}=2{\bf i}$, ${\bf {\it b}}=2({\bf i}+{\bf j})$ and ${\bf{\it c}}=2({\bf i}+{\bf j}+{\bf k})$. It is easily seen that the above three vectors make a primitive cell for a cubic lattice.

\begin{figure} [ht] 
\begin{center} 
\includegraphics[scale=0.5, angle=0]{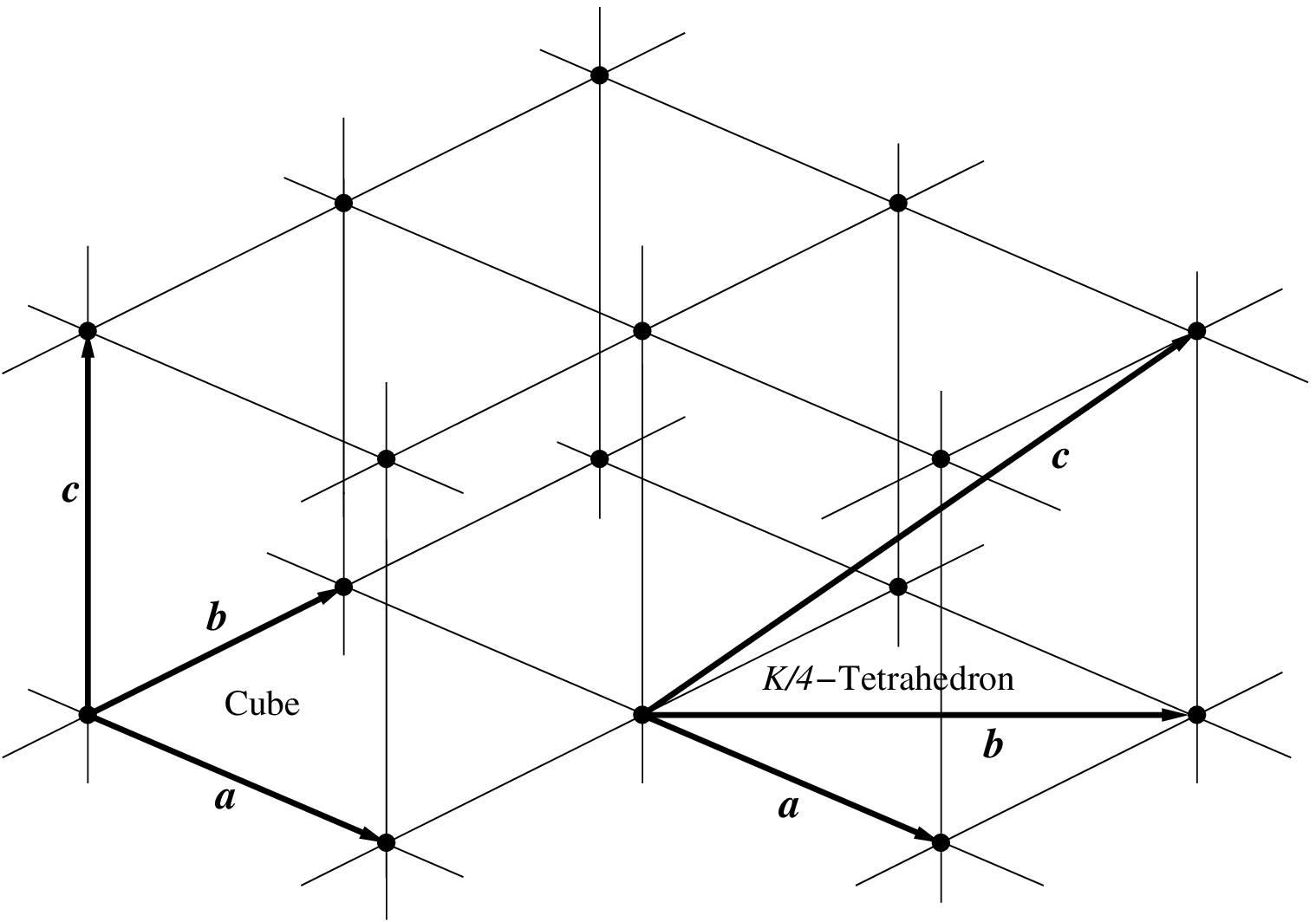}
\hspace{0.2in}
\includegraphics[scale=0.5, angle=0]{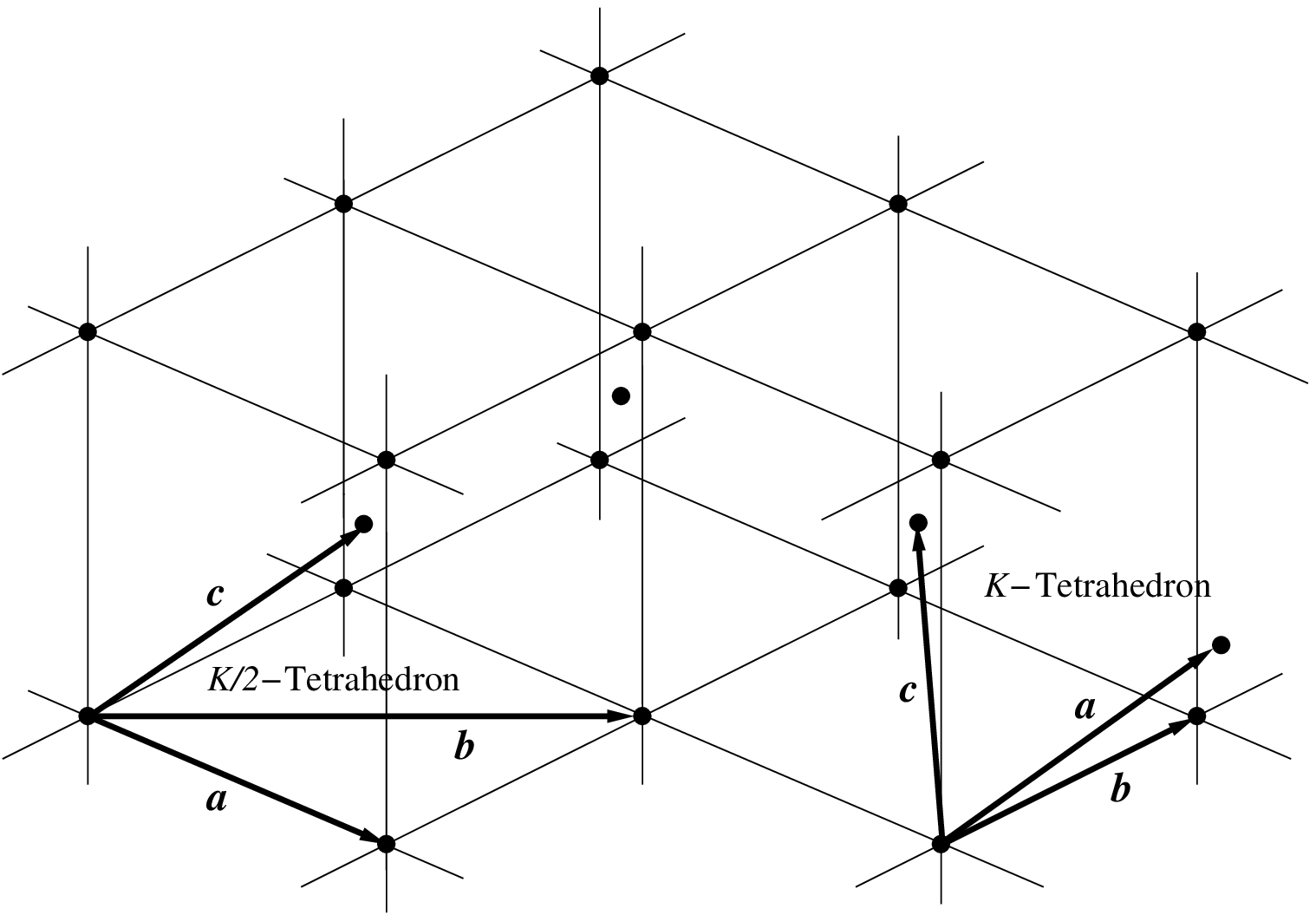}
\caption{Lattice vectors for geometric constructions of $3$-dimensional domains. Left: simple-cubic; Right: body-centered-cubic; Lattice constant for the cubic lattice is set to $2$ units, whereas, the same for the body-centered is set to $4$ units.}
\label{3dlattice}
\end{center}
\end{figure}

Fig.\ref{3dlattice} shows all four sets of lattice vectors for $3$-dimensional integrable domains. Again we note that the two cases - the cube and the $K/4$-tetrahedron are generated from a simple cubic primitive cell basis vectors. The remaining two, i.e., the $K$- and the $K/2$-tetrahedra are generated from a body-centered cubic primitive cell basis vectors.
The other integrable domain would come from the primitive orthorhombic system. This would be a trivial extension of the cubic system with mutually orthogonal basis vectors of different length as generators. One should further note that in each case the three basis vectors are along the three edges of the corresponding integrable domain. This could be easily seen by comparing the Figs.\ref{somertetra} and \ref{3dlattice}.

We summarize the results of $3$-dimensional integrable domains in table \ref{3dsum}. For all the cases, twice the amplitude vector is
${\bf V}=l{\bf {\it a}}+m{\bf {\it b}}+n{\bf{\it c}}$. The integers $l, m$ and $n$ take values $1,2,3, \cdots$. Further, we show that in all the cases the quantum energy is proportional to the square of the amplitude, ${\cal A}(l,m,n)= \left[A(l,m,n)\right]^2=({1\over 2}|{\bf V}|)^2$, i.e.,
\be
E_{l,m,n}\propto {\cal A}(l,m,n);\quad l,m,n=1,2,3,\cdots.
\ee
\begin{table}
\caption{Lattice vectors and amplitude-squares for 3D integrable billiards.}
\begin{center}
\scalebox{0.95}{\begin{tabular}[ht]{*{4}{c}}\toprule\midrule
Shape & Lattice vectors &
Amplitude$^2$=$|{\bf V}|^2/4$\\
\toprule
Cube & ${\bf {\it a}}=2{\bf i}$; ${\bf{\it b}}=2{\bf j}$; ${\bf{\it c}}=2{\bf k}$ & $l^2+m^2+n^2$\\ \midrule
$K$-Tetrahedron & ${\bf {\it a}}=2({\bf i}+{\bf j}+{\bf k})$; ${\bf {\it b}}=4{\bf j}$; 
${\bf{\it c}}=2(-{\bf i}+{\bf j}+{\bf k})$ 
& $3l^2+4m^2+3n^2+4lm+2ln+4mn$\\ \midrule
$K/2$-Tetrahedron & ${\bf {\it a}}=4{\bf i}$; ${\bf {\it b}}=4({\bf
 i}+{\bf j})$; ${\bf {\it c}}=2({\bf i}+{\bf j}+{\bf k})$ &
$4l^2+8m^2+3n^2+ 8lm+4ln+8mn $ \\ \midrule
$K/4$-Tetrahedron & ${\bf {\it a}}=2{\bf i}$; ${\bf {\it b}}=2({\bf i}+{\bf
 j})$; ${\bf {\it c}}=2({\bf i}+{\bf j}+{\bf k})$ & $l^2+2m^2+3n^2+ 2lm+2ln+4mn $ \\ \midrule
\bottomrule
\end{tabular}}
\end{center}
\label{3dsum}
\end{table}
Numerical computation confirms that (\ref{k2ftetra}) and (\ref{k4ftetra}), indeed provide the eigenspectra of the Helmholtz equation in the $K/2$- and $K/4$- tetrahedral domains respectively with Dirichlet boundary condition (DBC). Comparison of numerical (NV) and analytical (AV) results are shown in table \ref{numana} for the first 40 states. Moreover, it is interesting to note that the same expressions are also valid for the spectra of the Helmholtz equation with Neumann boundary condition (NBC) in the corresponding tetrahedral domains. In this case, $l,m$ and $n$ are free to take the value 0, except for the case $l=m=n=0$ like the $2$-dimensional cases described in the earlier sections. Table \ref{numana} also shows the comparison between the numerical and the analytical results for Neumann boundary condition. We find that the results for all the cases agree remarkably well.

\section{Conclusions and summary}
In this paper, we have established an interesting one-to-one mapping between the squared amplitude of a periodic trajectory of a classical particle trapped in a class of 2-dimensional billiards and the energy of a quantum particle confined to move in an identical region with infinitely high potential wall on the boundary. The quantum energy-spectra and the wave solution are obtained by solving Helmholtz equation in the same domain and imposing Dirichlet boundary condition. We have also shown that the classical periodic trajectories which are incident normally on any of the boundaries, correspond to wave solutions of the Helmholtz equation in identical domain but with Neumann boundary condition. The exact solutions for a particular domain are dictated by the geometry of the boundary and symmetries of the domain. An integrable domain should satisfy the reflective tiling condition - it should be possible to cover the whole of $\mathbb{E}^2$ with congruent tiles without leaving any empty space. Tiling of $\mathbb{E}^{2}$ immediately enables us to build up a lattice structure with the domain as a fundamental unit. Since there is no potential inside the domain, the classical particle trajectory is a straight line on this lattice. Any straight vector on this lattice which begins and ends on lattice points has the potential of becoming a periodic trajectory. Reflective tiling is the key to the geometric construction and it is exploited to fold a lattice vector into a fundamental domain to identify it as a classical trajectory.
 
We have further extended the method to billiards in three dimensions for which exact analytical results are hardly available. Extending the geometric construction, we identify and extract the exact energy spectra of two new tetrahedral domains, $K/2$ and $K/4$, which we believe are integrable. We test the validity of our approach by comparing our results with those obtained numerically. It would be an interesting avenue of investigation to explore whether these $3$-dimensional domains exhaust the possibility of integrability in three
dimensions or one may have other polyhedral domains for which exact solutions exist with Dirichlet or Neumann boundary conditions - when we say other polyhedral domains, we mean the genuine $3$-dimensional domains and not the prism-type polyhedra where the $2$-dimensional base is integrable and there is a third orthogonal direction reducing the problem to two dimensions effectively. Like the cases in two dimensions, we believe that a unified reciprocal lattice approach could be taken for the integrable domains in three dimensions where the moduli constructed from the reciprocal lattice vectors would be isospectral with the spectrum of corresponding amplitudes of the daughter lattice vectors.

Our geometrical constructions are intimately related with the primitive cells of Bravais lattices in two and three dimensions. This may provide another possible direction for future studies since there are fourteen Bravais lattices in three dimensions and we have used only two of them, viz. simple cubic and body-centered cubic, for our geometric construction.


\section*{Acknowledgments}
The authors would like to thank Prasun Dutta for fruitful discussions during various stages of the work.

\newcommand{\ra}[1]{\renewcommand{\arraystretch}{#1}}
\renewcommand\tabcolsep{6pt}
\begin{table}
\caption{Comparison of analytical and numerical values for 3D integrable
 domains. For DBC analytical values, $l,m,n=1,2,3,\cdots$ and for the NBC, 
$l,m,n=0,1,2\cdots$ except $l=m=n=0$.}
\ra{1.04}
\begin{center}
\scalebox{0.75}{\begin{tabular}{*{20}{c}}\toprule\midrule
\multirow{3}{*}{} & \multicolumn{5}{c}{$K$-Tetrahedron} & &&
\multicolumn{5}{c}{$K/2$-Tetrahedron} && &\multicolumn{5}{c}{$K/4$-Tetrahedron}
\\ \cmidrule{2-6} \cmidrule{9-13}\cmidrule{16-20}
& \multicolumn{5}{c}{AV=$3l^2+4m^2+3n^2$} &&& \multicolumn{5}{c}{AV=$4l^2+8m^2+3n^2$}
&& &\multicolumn{5}{c}{AV=4($l^2+2m^2+3n^2$}\\ 
& \multicolumn{5}{c}{$+4lm+2ln+4mn$} &&& \multicolumn{5}{c}{$+8lm+4ln+8mn$} && &\multicolumn{5}{c}{$2lm+2ln+4mn$)}\\ 
\cmidrule{2-6} \cmidrule{9-13}\cmidrule{16-20}
\multirow{1}{*}{Sl.} & \multicolumn{2}{c}{DBC} && \multicolumn{2}{c}{NBC} & &&\multicolumn{2}{c}{DBC} && \multicolumn{2}{c}{NBC} & &&\multicolumn{2}{c}{DBC} && \multicolumn{2}{c}{NBC} \\ \cmidrule{2-3} \cmidrule{5-6} \cmidrule{9-10} \cmidrule{12-13}\cmidrule{16-17} \cmidrule{19-20}
\multirow{1}{*}{No.} & AV & NV && AV & NV &&& AV & NV && AV & NV &&& AV & NV && AV & NV \\ \cmidrule{1-6} \cmidrule{9-13}\cmidrule{16-20}
 1 & 20 & 20.00 && 3 & 3.00 & && 35 & 35.00 && 3 & 3.00 & &&56 & 56.00 && 4 & 4.00 \\
 2 & 35 & 35.00 && 3 & 3.00 & &&56 & 56.01 && 4 & 4.00 & &&84 & 84.01 && 8 & 8.00 \\
 3 & 35 & 35.01 && 4 & 4.00 & && 59 & 59.01 && 8 & 8.00 && &104 & 104.03 && 12 & 12.00 \\
 4 & 40 & 40.01 && 8 & 8.00 & &&75 & 75.03 && 11 & 11.00 && &116 & 116.05 && 16 & 16.00 \\
 5 & 52 & 52.02 && 11 & 11.00 &&& 83 & 83.05 && 12 & 12.00 &&& 120 & 120.05 && 20 & 20.00 \\
 6 & 56 & 56.02 && 11 & 11.00 && &84 & 84.05 && 16 & 16.00 && &140 & 140.10 && 24 & 24.00 \\
 7 & 56 & 56.03 && 12 & 12.00 && &91 & 91.07 && 19 & 19.00 && &152 & 152.13 && 32 & 32.00 \\
 8 & 59 & 59.03 && 12 & 12.00 && &104 & 104.10 && 20 & 20.00 &&& 164 & 164.15 && 36 & 36.00 \\
 9 & 59 & 59.03 && 16 & 16.00 && &107 & 107.11 && 24 & 24.00 && &168 & 168.18 && 36 & 36.00 \\
 10 & 68 & 68.05 && 19 & 19.00 && &115 & 115.14 && 27 & 27.00 && &180 & 180.22 && 40 & 40.00 \\
 11 & 75 & 75.07 && 19 & 19.00 && &116 & 116.14 && 27 & 27.00 && &184 & 184.23 && 44 & 44.00 \\
 12 & 75 & 75.07 && 20 & 20.00 && &120 & 120.16 && 32 & 32.00 && &196 & 196.29 && 48 & 48.00 \\
 13 & 80 & 80.08 && 24 & 24.00 && &131 & 131.21 && 35 & 35.00 && &200 & 200.31 && 52 & 52.01 \\
 14 & 83 & 83.09 && 24 & 24.00 && &131 & 131.22 && 36 & 36.00 && &212 & 212.36 && 56 & 56.01 \\
 15 & 83 & 83.09 && 27 & 27.00 && &139 & 139.25 && 36 & 36.01 && &216 & 216.34 && 64 & 64.01 \\
 16 & 84 & 84.09 && 27 & 27.00 && &140 & 140.25 && 40 & 40.01 && &224 & 224.42 && 68 & 68.01 \\
 17 & 84 & 84.10 && 27 & 27.01 && &147 & 147.30 && 43 & 43.01 && &236 & 236.47 && 68 & 68.01 \\
 18 & 91 & 91.12 && 27 & 27.01 && &152 & 152.32 && 44 & 44.01 && &244 & 244.56 && 72 & 72.02 \\
 19 & 91 & 91.12 && 32 & 32.01 && &155 & 155.33 && 48 & 48.01 && &248 & 248.55 && 72 & 72.02 \\
 20 & 100 & 100.16 && 35 & 35.01 &&& 155 & 155.36 && 51 & 51.01 &&& 248 & 248.59 && 76 & 76.02 \\
 21 & 104 & 104.17 && 35 & 35.01 && &164 & 164.40 && 51 & 51.01 &&& 260 & 260.70 && 80 & 80.02 \\
 22 & 104 & 104.18 && 36 & 36.01 && &168 & 168.43 && 52 & 52.02 &&& 264 & 264.69 && 84 & 84.03 \\
 23 & 104 & 104.19 && 36 & 36.01 && &171 & 171.46 && 56 & 56.02 &&& 276 & 276.72 && 88 & 88.03 \\
 24 & 107 & 107.19 && 36 & 36.01 && &179 & 179.53 && 59 & 59.02 &&& 276 & 276.83 && 96 & 96.04 \\
 25 & 107 & 107.20 && 40 & 40.02 && &179 & 179.56 && 59 & 59.02 &&& 280 & 280.85 && 100 & 100.04 \\
 26 & 115 & 115.23 && 43 & 43.02 && &180 & 180.54 && 64 & 64.03 &&& 296 & 296.96 && 100 & 100.04 \\
 27 & 115 & 115.24 && 43 & 43.02 && &184 & 184.57 && 67 & 67.03 &&& 296 & 297.04 && 104 & 104.05 \\
 28 & 116 & 116.23 && 44 & 44.02 && &195 & 195.67 && 68 & 68.03 &&& 300 & 301.03 && 104 & 104.05 \\
 29 & 116 & 116.24 && 44 & 44.02 && &195 & 195.70 && 68 & 68.04 &&& 308 & 309.11 && 108 & 108.05 \\
 30 & 116 & 116.25 && 48 & 48.03 && &196 & 196.68 && 72 & 72.04 &&& 308 & 309.19 && 108 & 108.06 \\
 31 & 120 & 120.27 && 48 & 48.03 && &200 & 200.74 && 72 & 72.04 &&& 312 & 313.14 && 116 & 116.07 \\
 32 & 120 & 120.29 && 51 & 51.03 && &203 & 203.77 && 75 & 75.04 &&& 324 & 325.23 && 116 & 116.07 \\
 33 & 131 & 131.35 && 51 & 51.03 && &203 & 203.80 && 75 & 75.05 &&& 332 & 333.39 && 120 & 120.08 \\
 34 & 131 & 131.35 && 51 & 51.03 && &211 & 211.87 && 76 & 76.05 &&& 336 & 337.43 && 128 & 128.09 \\
 35 & 131 & 131.36 && 51 & 51.03 && &212 & 212.88 && 80 & 80.05 &&& 344 & 345.40 && 132 & 132.10 \\
 36 & 131 & 131.38 && 52 & 52.04 && &216 & 216.91 && 83 & 83.06 &&& 344 & 345.55 && 132 & 132.10 \\
 37 & 136 & 136.40 && 56 & 56.04 && &219 & 219.97 && 83 & 83.06 &&& 356 & 357.67 && 136 & 136.10 \\
 38 & 139 & 139.41 && 56 & 56.05 && &224 & 225.04 && 84 & 84.07 &&& 356 & 357.80 && 136 & 136.11 \\
 39 & 139 & 139.42 && 59 & 59.05 && &227 & 228.05 && 88 & 88.07 &&& 360 & 361.77 && 140 & 140.13 \\
 40 & 140 & 140.41 && 59 & 59.05 && &227 & 228.12 && 91 & 91.08 &&& 360 & 361.78 && 144 & 144.12 \\
\midrule
\bottomrule
\multicolumn{5}{c}{AV=Analytical Value,}&\multicolumn{5}{c}{NV=Numerical
 Value,}&\multicolumn{5}{c}{DBC=Dirichlet BC,}&\multicolumn{5}{c}{NBC=Neumann
 BC}
\end{tabular}}
\end{center}
\label{numana}
\end{table}

\end{document}